\documentclass[11pt,a4paper]{article}
\usepackage{jheppub,longtable,feynmp-auto,slashed,ulem,xpatch}

\newcommand{\bea}{\begin{eqnarray}}
\newcommand{\eea}{\end{eqnarray}}
\newcommand{\be}{\begin{equation}}
\newcommand{\ee}{\end{equation}}
\def\bsp#1\esp{\begin{split}#1\end{split}}
\def\bpm{\begin{pmatrix}}
\def\epm{\end{pmatrix}}

\def\W{{\tilde W}}
\def\Q{{\tilde Q}}
\def\g{{\tilde g}}
\def\B{{\tilde B}}
\def\tchi{{\tilde \chi}}
\def\Emiss{{\slashed{E}_T}}

\newcommand{\ie}{{\it i.e.}}
\newcommand{\eg}{{\it e.g.}}

\xpatchcmd{\sout}
  {\bgroup}
  {\bgroup}
  {}{}

\begin{document}
%
\preprint{{\tiny CUMQ/HEP 191,HIP-2017-01/TH,HRI-RECAPP-2017-001} }

\title{Resonant slepton production and right sneutrino dark matter in left-right
   supersymmetry}
\author[1]{Mariana Frank}
\author[2,3,4]{\!\!, Benjamin Fuks}
\author[5]{\!\!, Katri Huitu}
\author[6]{\!\!, Santosh Kumar Rai}
\author[5]{\! and Harri Waltari}

\affiliation[1]{Department of Physics, Concordia University, 7141 Sherbrooke St.
  West, Montreal, Quebec, CANADA H4B 1R6}
\affiliation[2]{Sorbonne Universit\'es, UPMC Univ.~Paris 06, UMR 7589, LPTHE,
   F-75005 Paris, France}
\affiliation[3]{CNRS, UMR 7589, LPTHE, F-75005 Paris, France}
\affiliation[4]{Institut Universitaire de France, 103 boulevard Saint-Michel,
  75005 Paris, France}
\affiliation[5]{Department of Physics and Helsinki Institute of Physics, University of Helsinki,
   P.O. Box 64 (Gustaf H\"allstr\"omin katu 2), FI-00014 University of Helsinki,
   Finland,}
\affiliation[6]{Regional Centre for Accelerator-based Particle Physics,
   Harish-Chandra Research Institute, HBNI, Chhatnag Road, Jhusi, Allahabad
   211019, India.}

\emailAdd{mariana.frank@concordia.ca}
\emailAdd{fuks@lpthe.jussieu.fr}
\emailAdd{katri.huitu@helsinki.fi}
\emailAdd{skrai@hri.res.in}
\emailAdd{harri.waltari@helsinki.fi}

\date{\today}

\abstract{Right-handed sneutrinos are natural components of left-right symmetric supersymmetric 
models where the gauge sector is extended to include right-handed weak interactions. 
Unlike in other models where right-handed sneutrinos are gauge singlets, here the right 
sneutrino is part of a doublet and could be a dark matter candidate whose
annihilation proceeds via gauge interactions.
We investigate this possibility, and find that relic density, low-energy
observable and direct supersymmetry search constraints can be
satisfied when the lightest supersymmetric particle is a right-handed sneutrino.
We introduce benchmarks for left-right supersymmetric realizations where either
a sneutrino or a neutralino is the lightest superpartner.
We then study the LHC signals arising
through resonant right-handed slepton production via a
$W_R$ gauge-boson exchange 
that lead to f\mbox{}inal states enriched in leptons, additionally containing a
large amount of missing transverse momentum, and featuring a low jet
multiplicity. We find that such a resonant production would 
boost the chances of discovering these weakly interacting supersymmetric
particles for a mass range extending beyond 1~TeV already with a
luminosity of 100 fb$^{-1}$. Finally, we compare sneutrino versus 
neutralino scenarios, and comment on differences with other sneutrino dark
matter models.}

\keywords{Supersymmetry phenomenology}
\maketitle

\section{Introduction}
\label{sec:intro}
While the LHC Run 1 has established the existence of a Higgs boson with properties 
consistent with those of the Standard Model (SM) 
one \cite{Aad:2012tfa,Chatrchyan:2012xdj} and found no other new particles, the 
outstanding theoretical problems of the  SM remain unresolved. In addition, the existence of 
dark matter (DM) weighs heavily on the list of experimentally observed  but theoretically 
unexplained problems. One could argue that supersymmetry (SUSY), which provides the 
best motivated candidate for DM  in the lightest supersymmetric particle (LSP),
still stands as the best candidate of physics beyond the SM. 
Unfortunately no signals of supersymmetry have been observed yet, discrediting its simplest 
incarnation, the constrained minimal supersymmetric standard model (MSSM). In
the latter configuration, direct 
collider bounds push the masses of the strongly interacting supersymmetric partners (gluino 
and squarks) to be larger than about 1~TeV, which also affects sleptons (their masses being derived from 
the same universal scalar mass $m_0$ as the squarks) and electroweak gauginos (their masses depending on
the same universal mass $m_{1/2}$ as the gluino). In addition, the higgsino mass is 
under pressure from direct searches, leading to a situation in which neutralino DM (either 
bino- or higgsino-dominated) is in jeopardy. An alternative solution would be to abandon the 
minimal spectrum of the MSSM and introduce additional symmetries and/or
particles which could serve as DM candidates, keeping in mind that SUSY mass 
limits depend critically on the nature and on the mass of the LSP. 

A viable DM alternative has been provided by the sneutrino, which requires the
MSSM to be augmented by at least one right-handed (RH) neutrino
superfield~\cite{ArkaniHamed:2000bq,Borzumati:2000mc}.
The left-handed (LH) sneutrino of the MSSM has a non-zero hypercharge
and is thus excluded as a DM candidate. Its coupling to the $Z$ boson indeed
causes it to annihilate too much in the early Universe, yielding a relic density
much lower than the one measured by the WMAP and Planck
satellites~\cite{Hinshaw:2012aka,Ade:2015xua}. In addition, the presence of a RH
neutrino superfield helps in providing a mechanism for generating light neutrino
masses which is otherwise absent in the MSSM.

The phenomenology of the MSSM with RH sneutrinos has been investigated in
detail, including implications for direct and indirect DM detection, signals at
the LHC, and restrictions on the model parameter space~\cite{Belanger:2010cd}.
One could also abandon the MSSM formalism and look for extended supersymmetric models, which cure some of the problems that the MSSM inherits from the SM and where the sneutrino emerges as a natural DM candidate \cite{Chen:2015yuz}. However, most analyses have been performed in the case where the sneutrino is a gauge
singlet, which results in LH and RH sneutrino mixings through additional Yukawa and trilinear couplings independent of the LH sneutrinos.

No full investigation exists for the case where 
RH sneutrinos belong to doublets and the theory possesses a symmetry linking the
interactions of the LH and RH fields.\footnote{In Ref.~\cite{An:2011uq}, the authors consider a different version of left-right models, where an additional (s)neutrino, singlet under SU(2)$_L$ and SU(2)$_R$, is added. The seesaw mechanism induces a mixing between the left-handed, right-handed, and singlet sneutrinos which can yield the right relic density through $Z$-channel annihilation.}
  This is the case of the supersymmetric
left-right model (LRSUSY) which we propose to investigate here. The LRSUSY model
includes three generations of RH neutrino superfields (as parts of RH lepton doublets), and
a seesaw as the mechanism for generating the neutrino
masses~\cite{Mohapatra:1979ia} emerges from choosing a triplet representation
for the Higgs field responsible for the breaking of the left-right symmetry. The RH neutrino partners are the RH sneutrinos, one of which could be the LSP and thus be a
DM candidate. We expect the collider signatures of this model to differ from the cases in which the RH sneutrino is a singlet, as now the sneutrino can couple
differently, through gauge interactions.

Left-right supersymmetric models  have been explored before 
\cite{Francis:1990pi}. They have several attractive features, such that the fact that they 
account for neutrino masses, parity violation,  disallow explicit R-parity violation, offer
a solution to the strong and weak $CP$ violation problems without requiring to
introduce an axion
\cite{Mohapatra:1995xd}, and explain the absence of excessive SUSY $CP$
violation. Left-right symmetry is moreover favored by many
extra-dimensional models and many gauge unification scenarios, such as $SO(10)$.
We shall explore here left-right supersymmetric realizations which, while
protecting against spontaneous $R$-parity violation, introduces only one extra
singlet superfield. A specific related aspect that has been overlooked so far
concerns the possibility of resonantly producing supersymmetric particles at colliders,
in particular due to the presence of extra gauge and Higgs bosons.
In this context, we explore dark matter constraints and the expected resonant collider signals, paying particular attention to distinguishing features.

Our work is organized as follows. In Sec.~\ref{sec:lrsusy} we give a description
of the model, with a particular emphasis on the slepton, sneutrino, chargino and
neutralino sectors. We impose current experimental constraints on the model and
define our benchmarks in Sec.~\ref{sec:constraints}. We then discuss the
phenomenology for cases where the dark matter candidate is either a sneutrino or
a neutralino in Sec.~\ref{sec:dm} and analyze the characteristic signature of this model at the LHC in Sec.~\ref{sec:signals}. Finally, we summarize and conclude our analysis in Sec.~\ref{sec:conclusion}.

\section{A Left-Right Supersymmetric Model with $R$ parity conservation}
\label{sec:lrsusy}
\subsection{Model description}
There are several realizations of a left-right symmetry in supersymmetry. In its
general formulation, the model is based on the $SU(3)_c \otimes SU(2)_L \otimes
SU(2)_R \otimes U(1)_{B-L}$ gauge symmetry. The embedding as a gauge symmetry of
the only quantum number left ungauged in the SM, ${B-L}$ where $B$ and $L$
stand for the baryon and lepton numbers, is an additional attractive feature.
The model contains left and right fermion doublets, as well as two sets of gauge
bosons lying in the adjoint representation of the $SU(2)_L$ and $SU(2)_R$ groups
and a neutral gauge boson connected to $U(1)_{B-L}$. While $R$-parity, defined
as $R_P = (-1)^{3(B-L)+2s}$ (with $s$ being the spin of the particle), is
imposed in the MSSM to avoid dangerous baryon and lepton number violating
operators, explicit $R$-parity breaking is forbidden in LRSUSY by the model
symmetries. Extra Higgs fields must be introduced to spontaneously break the
LRSUSY symmetry group, of which $SU(2)_{R}$ Higgs triplets consist of the most
attractive option as they induce a seesaw mechanism for neutrino mass
generation~\cite{Mohapatra:1979ia}. $R$-parity may however not be conserved in
this setup, this discrete symmetry being broken spontaneously as the vacuum prefers a
solution in which the RH sneutrino gets a vacuum expectation value (VEV). Two 
scenarios have been proposed to remedy this situation. One possibility is to
introduce an extra singlet Higgs boson so that a stable $R$-parity-conserving
minimum is found once one-loop corrections are added to the
potential~\cite{Babu:2008ep,Frank:2014kma}, while a second option requires to
add two new Higgs triplets uncharged under the $B-L$ symmetry
$\Omega (1, 3, 1, 0)$ and $\Omega_c (1, 1, 3, 0)$
to break the left-right symmetry spontaneously while conserving
$R$-parity at tree-level~\cite{Aulakh:1997fq}. Here we adopt the former, as it is a more minimal realization, for which we present a short description below. 

The matter sector of our LRSUSY model contains quark and lepton doublet
superfields,
\be\bsp
    (Q_L)^{  i  } = \bpm u_L^{i }\\d_L^{i } \epm =
      \big({\bf 3}, {\bf 2},{\bf 1},\frac13\big) \ , 
 &\qquad 
    (Q_R)^{i} = \bpm d_{R}^i \\ -u_{R}^i   \epm=
      \big({\bf \bar 3}, {\bf 1},{\bf 2}^*,-\frac13\big) \ ,\\
    (L_L)^{ i} = \bpm \nu_L^i \\ \ell_L^i\epm =
     \big({\bf 1}, {\bf 2},{\bf 1},-1\big)\ ,
 & \qquad 
    (L_R)^{i} = \bpm \ell_{R}^i  \\ -\nu_{R}^i  \epm =
     \big({\bf 1}, {\bf 1},{\bf 2}^*,1\big)\ ,
\esp 
\label{eq:fieldcontent} \ee
where the respective representations under the $SU(3)_c \otimes SU(2)_L \otimes
SU(2)_R \otimes U(1)_{B-L}$ gauge symmetry have been indicated. The Higgs sector
is in contrast more complicated and features various superfields,
\be
\bsp
  \Phi_1 = \bpm \phi_1^+&\phi_1^{0 \prime}\\ \phi^{0}_1 &\phi_1^-\epm =
     \big( {\bf 1}, {\bf 2}, {\bf 2}^*,0\big) \ ,
  & \qquad 
    \Phi_2 = \bpm \varphi_{2}^+&\varphi_2^{0 }\\ \varphi_2^{0 \prime }&\varphi_2^- \epm  =
     \big( {\bf 1},  {\bf 2}, {\bf 2}^*,0\big)\ , \\
    \Delta_{1L} =
    \bpm
      \frac{\delta^{-}_{1L}}{\sqrt{2}} & \delta_{1L}^{0}\\
      \delta_{1L}^{--} & -\frac{\delta_{1L}^{-}}{\sqrt{2}}
    \epm = \big( {\bf 1}, {\bf 3}, {\bf 1}, -2\big)\ ,
   & \qquad \Delta_{2L} =
     \bpm
       \frac{\delta^{+}_{2L}}{\sqrt{2}} & \delta_{2L}^{++}\\
       \delta_{2L}^{0} & -\frac{\delta_{2L}^{+}}{\sqrt{2}}
     \epm = \big( {\bf 1},  {\bf 3}, {\bf 1} ,2\big) \ , \\
     \Delta_{1R} = 
      \bpm
        \frac{\delta_{1R}^-}{\sqrt{2}}  & \delta_{1R}^0\\
        \delta_{1R}^{--} & -\frac{\delta_{1R}^-}{\sqrt{2}} 
      \epm  = \big( {\bf 1},  {\bf 1}, {\bf 3},-2\big)\ , 
   & \qquad  \Delta_{2R}  = 
      \bpm
        \frac{\delta_{2R}^+}{\sqrt{2}} & \delta_{2R}^{++}\\
        \delta_{2R}^0 & -\frac{\delta_{2R}^+}{\sqrt{2}} 
      \epm = \big( {\bf 1},  {\bf 1}, {\bf 3} ,2\big) \ , \\
       \qquad
      &\hspace{-1.2cm} S = \big({\bf 1}, {\bf 1},{\bf 1},0\big) \ .
\esp
\label{eq:fieldcontent1}
\ee
The model superpotential is given by
\be\bsp
   & W =
   (Q_L)^{ T} Y_Q^1  \Phi_1 (Q_R) +
   (Q_L)^{ T} Y_Q^2  \Phi_2 (Q_R) +
   (L_L)^{T}Y_L^1   \Phi_1  (L_R) +
   (L_L)^{T} Y_L^2 \Phi_2 (L_R) \\
&\ +
   (L_L)^{T} Y_L^3 \Delta_{2L} (L_L) +
   (L_R)^{T} Y_L^4 \Delta_{1R} (L_R) 
   + S \Big[\lambda_L {\rm Tr}(\Delta_{1L}\cdot \Delta_{2L})
   + \lambda_R \,{\rm Tr}(\Delta_{1R} \cdot  \Delta_{2R} )\\
   &\ + \lambda_{3} {\rm Tr}(\Phi_{1}^{T}\tau_{2}\Phi_{2}
     \tau_{2})+\lambda_{4}{\rm Tr}(\Phi_{1}^{T}\tau_{2}\Phi_{1}\tau_{2})
   + \lambda_{5}{\rm Tr}(\Phi_{2}^{T}\tau_{2}\Phi_{2}\tau_{2})
   +\lambda_{S}S^{2}+\xi_{F}\Big]\ ,
\esp\label{eq:Wtrip}\ee
where generation indices are suppressed for clarity. Following the conventions
of Ref.~\cite{Alloul:2013fra}, the Yukawa couplings
$Y^j_Q$ and  $Y^j_L$ are $3\times 3$ matrices in flavour space, the $\lambda$
parameters denote various trilinear Higgs interactions (with $\tau_2$ being the
second Pauli matrix) and $\xi$ a linear singlet term. Explicit bilinear supersymmetric
Higgs mass terms are in principle allowed by the model symmetries, but we omit
them. The bilinear terms are nevertheless dynamically generated
when the scalar singlet field $S$ gets a vacuum expectation value.

The left- and right-handed matter superf\mbox{}ields and gauge sectors can be related
through the parity transformation \cite{Mohapatra:1995xd}. 
Since we may impose parity symmetry on the Lagrangian,
the parity violating $\tilde{G}^{\mu\nu}G_{\mu\nu}$-term is absent and the gluino mass
parameter is real. Left-right symmetry further imposes the Yukawa matrices to be Hermitian
and the same is true for the soft trilinear terms. The hermiticity of Yukawa matrices
also makes the VEVs of the MSSM-like bidoublet Higgses (denoted by $v_1$, $v_2$ in the following)
real, and hence the model can explain both the strong and SUSY $CP$
problems without introducing the axion as long as the parity breaking scale is not too large~\cite{Babu:2008ep,Mohapatra:1997su}.

The gauge symmetry is spontaneously broken once the Higgs fields acquire their
VEVs,
\be\bsp
   & \langle S \rangle = \frac{v_S}{\sqrt{2}} e^{i
    \alpha_S}\ ,
  \quad
  \langle \Phi_1 \rangle =  \bpm 
     \quad 0&  \frac{v_1^\prime}{\sqrt{2}}  e^{i \alpha_1} \quad  \\ 
      \frac{v_1}{\sqrt{2}}  &0 \epm \ , 
  \quad 
  \langle \Phi_2 \rangle = \bpm 
     \quad 0& \frac{v_2 }{\sqrt{2}} \quad \\ 
     \frac{v_2^\prime }{\sqrt{2}} e^{i \alpha_2} &0 \epm \ , \\
  &
    \langle \Delta_{1R} \rangle =  \bpm 0 & \frac{v_{1R}}{\sqrt{2}} \\ 0 & 0 \epm , \
  \quad
  \langle \Delta_{2R} \rangle =  \bpm 0 & 0 \\ \frac{v_{2R}}{\sqrt{2}} & 0 \epm \ ,
\esp
\ee
where we assume the LH triplets $\Delta_{1L}$, $\Delta_{2L}$ to be inert. This
is on one hand motivated by the constraints arising from the $\rho$
parameter and stems on the other hand from the radiative corrections to the
doubly-charged Higgs mass that must be significant enough to satisfy the current
experimental bounds. Both these prevent the LH triplet VEVs from being large.

The soft SUSY breaking Lagrangian reads
\be\bsp
& \mathcal{L}_{\rm soft} =  -\frac{1}{2}\left[M_{1}\B\B+M_{2L}\W_{L}^{a}\W_{La}+M_{2R}\W_{R}^{a}\W_{Ra}+M_{3}\g^{a}\g_{a}+{\rm h.c.} \right]
- m_{\Delta 1L}^{2}{\rm Tr}(\Delta_{1L}^{\dagger}\Delta_{1L})\\ &\
-m_{\Delta 2L}^{2}{\rm Tr}(\Delta_{2L}^{\dagger}\Delta_{2L})
- m_{\Delta 1R}^{2}{\rm Tr}(\Delta_{1R}^{\dagger}\Delta_{1R})-m_{\Delta 2R}^{2}{\rm Tr}(\Delta_{2R}^{\dagger}\Delta_{2R})
-m_{\Phi 1}^{2}{\rm Tr}(\Phi_{1}^{\dagger}\Phi_{1})\\&\
-m_{\Phi 2}^{2}{\rm Tr}(\Phi_{2}^{\dagger}\Phi_{2})
-m_{S}^{2}|S|^{2}+m_{\tilde{Q}_{L}}^{2}\Q^{\dagger}_{L}\Q_{L} -m_{\tilde{Q}_{R}}^{2}\Q^{\dagger}_{R}\Q_{R}- m_{\tilde L_L}^2( {\tilde L}_L^\dagger \tilde L_L)
-  m_{\tilde L_R}^2( {\tilde L}_R^\dagger \tilde L_R)\\&\
- \Big\{S [T_L {\rm Tr}(\Delta_{1L}\Delta_{2L})+T_R {\rm Tr}(\Delta_{1R}\Delta_{2R}) +T_{3}{\rm Tr} (\Phi_{1}^{T}\tau_{2} \Phi_{2}\tau_{2})+T_{4}{\rm Tr}(\Phi_{1}^{T}\tau_{2}\Phi_{1}\tau_{2})
\\&\ +T_{5}{\rm Tr}(\Phi_{2}^{T}\tau_{2}\Phi_{2}\tau_{2})+T_{S}S^{2}+\xi_{S}] + {\rm h.c.}\Big\} 
+ \Big\{ T_{Q}^{1}(\tilde Q_L)^{T} \Phi_1 (\tilde Q_R) + T_{Q}^{2} (\tilde Q_L)^{T} \Phi_2 (\tilde Q_R)\\ &\
 +T_{L}^{1}(\tilde L_L)^{T} \Phi_1 (\tilde L_R) + T_{L}^{2} (\tilde L_L)^{T}  \Phi_2 (\tilde L_R)  +T_L^{3}(\tilde L_L)^{T} \Delta_{2L} (\tilde L_L)+T_{L}^{4} (\tilde L_R)^{T} \Delta_{1R} (\tilde L_R) + {\rm h.c.}\Big\},
\esp\ee
and includes gaugino mass terms (first bracket), scalar mass terms (the $m^2$
terms) and trilinear scalar interactions whose strengths are given by the
$T$ couplings. For consistency with the superpotential, a linear $\xi$ term
has also been introduced.

The $v_{iR}$, $v_1$, $v_2$, $v^\prime_1$, $v^\prime_2$ and $v_S$ VEVs can be
chosen real and non-negative,  while the only complex phases which cannot be rotated away by
means of suitable gauge transformations  and field redefinitions are
denoted by the explicit angles $\alpha_1$, $\alpha_2$ and $\alpha_s$. However, the $CP$-violating $W^\pm_L-W^\pm_R$ mixing is proportional 
to $v_1 v_1^\prime e^{i \alpha_1}$ and $v_2 v_2^\prime e^{i
\alpha_2}$, and is constrained to be small by $K^0-{\bar K}^0$ mixing data.
To reduce the dimensionality of the parameter space, we therefore assume the hierarchy
\be \label{eq:vevhier}
  v_S,  v_{1R}, v_{2R} \gg  v_2, v_1, v_1^\prime, v_2^\prime \qquad\text{and}\qquad
 v_1^\prime  = v_2^\prime = \alpha_1 = \alpha_2 = \alpha_{S} \approx 0 \ .
\ee
This choice originates from the existing constraints on the $SU(2)_R$
gauge bosons that impose the RH VEVs to be large. In the supersymmetric limit,
the $F$-terms and $D$-terms vanish, when $\lambda_R v_{1R} v_{2R}=\xi_F$ and $v_{1R}=v_{2R}$~\cite{Babu:2008ep}.
On the other hand, the singlet VEV $v_S$ is induced by the SUSY-breaking linear
term $\xi_S$ so that its natural scale is the supersymmetry-breaking scale. We
finally realize an ad-hoc hierarchy $v_1, v_2 \gg v_1^\prime$, $v_2^\prime \approx 0$
by setting $\lambda_4$ and $\lambda_5$ and the corresponding SUSY-breaking 
parameters small. This is a convenient setup where one, for instance, avoids
potentially large flavour-changing neutral currents. For further references, we 
def\mbox{}ine $\tan \beta=v_2/v_1$ and $\tan \beta_R=v_{2R}/v_{1R}$.

The $D$-term contribution to the scalar potential (neglecting the squark pieces)
is given by
\be\bsp
V_D =&\ \sum_{i}\left[ \frac{g_{L}^{2}}{8}\left|{\rm Tr}(2\Delta_{1L}^{\dagger}\tau_{i}
\Delta_{1L}+2\Delta_{2L}^{\dagger}\tau_{i}\Delta_{2L}+\Phi_{a}\tau_{i}^{T}
\Phi_{b}^{\dagger}) +\tilde L^\dagger_L \tau_i \tilde L_L \right|^2\right. \\
&\qquad \left. +\frac{g_{R}^{2}}{8}\left|{\rm Tr}(2\Delta_{1R}^{\dagger}\tau_{i}
\Delta_{1R}+2\Delta_{2R}^{\dagger}\tau_{i}\Delta_{2R}+\Phi_{a}^{\dagger} \tau_{i}^{T}
\Phi_{b})+ \tilde L^\dagger_R \tau_i \tilde L_R \right|^2 \right]\\ &\qquad
+\frac{g_{B-L}^{2}}{2}\left[ {\rm
Tr}(-\Delta_{1L}^{\dagger}\Delta_{1L}+\Delta_{2L}^{\dagger}\Delta_{2L}-\Delta_{1R}
^{\dagger}\Delta_{1R}+\Delta_{2R}^{\dagger}\Delta_{2R} )-\tilde L^\dagger_L \tilde L_L +\tilde L^\dagger_R \tilde L_R \right]^{2}\ ,
\esp\ee
which yields, when expanded around the minimum of the potential, to a coupling between the SM-like Higgs boson and the imaginary parts of RH sneutrino fields.
Such a coupling is given, when the small neutrino Yukawa couplings are
neglected, by
\begin{equation}\label{eq:snucoupling}
\lambda_{h\tilde{\nu}_{RI}\tilde{\nu}_{RI}}=\frac{1}{4}g_{R}^{2}v\sin(\alpha+\beta)\simeq-\frac{1}{4}g_{R}^{2}v\cos 2\beta,
\end{equation}
where $\alpha$ stands for the mixing angle between the $\phi_1^0$ and
$\varphi_2^0$ fields and where the approximated form holds in the alignment
limit.
This coupling is essential when computing DM annihilation rates in the case of a
RH sneutrino LSP. At moderate or large values of $\tan \beta$, $\cos 2\beta \simeq -1$ so that $\lambda_{h\tilde{\nu}_{RI}\tilde{\nu}_{RI}}$ is nearly
independent of any free parameter, in particular if we assume $g_R\approx g_L$.

Minimizing the Higgs potential and solving the
$$\frac{\partial V}{\partial v_1}=  \frac{\partial V}{\partial
v_2}=\frac{\partial V}{\partial v_{1R}}=\frac{\partial V}{\partial v_{2R}}=\frac {\partial V}{\partial v_{S}}=0,$$
system of equations,
we derive the masses and compositions of the various Higgs bosons.
The correct minimum of the potential can however only be evaluated once the
one-loop Coleman-Weinberg effective potential
\begin{equation}
V_{\rm eff}^{\rm 1-loop}= \frac{1}{16 \pi^2} \sum_i(-1)^{2s} (2s+1)M_i^4 \left
[\ln \left (\frac{M_i^2}{\mu^2}\right)-\frac32 \right ].
\end{equation}
is included. Without this correction, the minimum would indeed be not phenomenologically acceptable and correspond to the
charge-breaking configuration
\begin{equation}
\langle \Delta_{1R}\rangle =\frac{1}{\sqrt{2}}
\left(
\begin{array}{cc}
0 & v_{1R}\\
v_{1R} & 0
\end{array}
\right),~~~~
\langle \Delta_{2R}\rangle =\frac{1}{\sqrt{2}}\left(
\begin{array}{cc}
0 & v_{2R}\\
v_{2R} & 0
\end{array}
\right).
\end{equation}  
We refer to and use the results of the recent extensive analysis of Ref.~\cite{Frank:2014kma} for the
calculation of the masses and mixing pattern of the Higgs sector, and focus
in the following subsections on the slepton, sneutrino, chargino and neutralino
sectors more relevant for this work in which we wish
to highlight the possibility for the sneutrino to be the LSP. It is nonetheless
equally interesting to look into the phenomenology of the other
sectors of the model when a sneutrino LSP is featured. This is left for future work.

\subsection{Charged sleptons and sneutrinos}
\label{subsec:sleptons}
In the interaction basis $(\tilde{L}_L^{i}, \tilde{L}_R^{i})$, the
squared-mass matrix for the sleptons is given by
\begin{equation}\label{eq:sleptonmass}
{\cal M}_{L}^2= 
 \bpm m_{\tilde L_L}^2+m_{\ell}^{2}+D_{11} & (T_{L}^{3})_{ij}v\cos \beta+\mu_{\mathrm{eff}}m_{\ell} \tan \beta\\
  (T_{L}^{3})_{ij}v\cos\beta+\mu_{\mathrm{eff}}m_{\ell} \tan \beta & m_{\tilde L_R}^2+m_{\ell}^{2}+D_{22}
 \epm,
\end{equation}
where $\mu_{\mathrm{eff}}=\lambda_{3}v_{s}/\sqrt{2}$ and where the $D$-terms
read
\bea
D_{11}&=&-\frac{g_{L}^{2}}{8}v^{2}\cos 2\beta+g_{B-L}^{2}(v_{1R}^{2}-v_{2R}^{2})
\quad\text{and}\nonumber\\
D_{22}&=&\frac{g_{R}^{2}}{8}\left[2(v_{1R}^{2}-v_{2R}^{2})-v^{2}\cos 2\beta \right]-g_{B-L}^{2}(v_{1R}^{2}-v_{2R}^{2}).
\label{eq:Dii}\eea
We then extract the scalar and pseudoscalar sneutrino mixing matrices that are of the form
\begin{eqnarray}
{\cal M}_{\tilde \nu}^2= \left( \begin{array}{cc}
 M_{\tilde \nu_L\tilde \nu_L}^2   & M_{\tilde \nu_L \tilde \nu_R}^2 \\
      M_{\tilde \nu_R\tilde \nu_L}^2  & M_{\tilde \nu_R \tilde \nu_R}^2                            \end{array}
                  \right)\ .
\end{eqnarray}
In the scalar case, the mass matrix entries are
\be\bsp
M_{\tilde \nu_L\tilde \nu_L}^2 =&\ m_{\tilde L_L}^2+ D_{11}\ ,\\
M_{\tilde \nu_L\tilde \nu_R}^2 =&\ M_{\tilde \nu_R\tilde \nu_L}^2=
(T_{L}^{2}v -Y_{L}^{2}Y_{L}^{4}v_{1R})\sin \beta +Y_{L}^{2}\mu_{\mathrm{eff}}\frac{v\cos\beta}{\sqrt{2}}\ ,  \\
M_{\tilde \nu_R \tilde \nu_R}^2=&\
  m_{\tilde L_R}^2+D_{22}+2(Y_{L}^{4})^{2}v_{1R}^{2}-\sqrt{2}T_{L}^{4}v_{1R}+Y_{L}^{4}\lambda_{R}v_{S}v_{2R},\label{eq:sneutrino1}
\esp\ee
where $D_{11}$ and $D_{22}$ are given in Eq.~\eqref{eq:Dii}. The terms depending on the Yukawa
couplings $Y_{L}^{2}$ that should have been included within the diagonal blocks have
been neglected, as they need to be small to get viable neutrino masses.
Moreover, the RH-LH neutrino mixing term will turn to be small as well, unless
$T_{L}^{2}$ is large.
The pseudoscalar mass matrix entries are given by
\be\bsp
M_{\tilde \nu_{IL}\tilde \nu_{IR}}^2 =&\ M_{\tilde \nu_{IR}\tilde \nu_{IL}}^2=
(T_{L}^{2}v +Y_{L}^{2}Y_{L}^{4}v_{1R})\sin \beta +Y_{L}^{2}\mu_{\mathrm{eff}}\frac{v\cos\beta}{\sqrt{2}}\\
M_{\tilde \nu_{IR} \tilde \nu_{IR}}^2=&\
  m_{\tilde L_R}^2+D_{22}+2(Y_{L}^{4})^{2}v_{1R}^{2}+\sqrt{2}T_{L}^{4}v_{1R}-Y_{L}^{4}\lambda_{R}v_{S}v_{2R}\label{eq:imsnumass}
\esp\ee
with $M_{\tilde{\nu}_{IL}\tilde{\nu}_{IL}}^{2}$ being identical to $M_{\tilde{\nu}_{L}\tilde{\nu}_{L}}^{2}$.
Adopting large values for $\lambda_{R}$ and a choice of positive parameters
implies that the last term of Eq.~\eqref{eq:imsnumass} will drive the sneutrino
masses. One of the pseudoscalar states, with a flavour aligned with the largest
element in the $Y_{L}^{4}$ matrix, will be the LSP unless the corresponding soft
supersymmetry breaking mass term is signif\mbox{}icantly larger than the 
other terms.

\subsection{Charginos and neutralinos}
\label{subsec:charginos}
We refer to Ref.~\cite{Alloul:2013fra} for detailed information on the chargino
and neutralino sector of the model. We recall below the corresponding mass
matrices that will be useful for the design of the benchmark scenarios relevant
for this work.

The model has six singly-charged charginos whose associated mass matrix can be
written in the $(\tilde{\Delta}_{L}^{\pm}, \tilde{\Delta}_{R}^{\pm},
\tilde{\Phi}_{1}^{\pm}, \tilde{\Phi}_{2}^{\pm}, \W_{L}^{\pm}, \W_{R}^{\pm})$
basis as
\begin{equation}
\label{eq:mchargino}
M_{\tchi^{\pm}}=
\bpm
\lambda_{L}v_{s}/\sqrt{2} & 0 & 0 & 0 & 0 & 0\\
0 & \lambda_{R} v_{s}/\sqrt{2} & 0 & 0 & 0 & -g_{R}v_{1R}\\
0 & 0 & 0 & \mu_{\mathrm{eff}} & g_{L}v_{u}/\sqrt{2} & 0\\
0 & 0 & \mu_{\mathrm{eff}} & 0 & 0 & -g_{R}v_{d}/\sqrt{2}\\
0 & 0 & 0 & g_{L}v_{d}/\sqrt{2} & M_{2L} & 0\\
0 & g_{R}v_{2R} & -g_{R}v_{u}/\sqrt{2} & 0 & 0 & M_{2R}\\
\epm\ ,
\end{equation}
where $v_{u}=v\sin \beta$, $v_{d}=v\cos \beta$.

Although the particle spectrum contains twelve neutralinos, the corresponding
mass matrix can be arranged into three block-diagonal pieces when the LH
triplet and two neutral bidoublet Higgs bosons are inert. The first two blocks
are expressed, in the $(\tilde{\delta}_{1L}, \tilde{\delta}_{2L})$ and
$(\tilde{\phi}_{2}, \tilde{\varphi}_{1})$ bases, as
\begin{equation}
M_{\tilde{\chi}_{\delta}}=
\begin{pmatrix}
0 & \mu_{L}\\
\mu_{L} & 0
\end{pmatrix}\qquad\text{and}\qquad
M_{\tilde{\chi}_{\Phi}}=
\begin{pmatrix}
0 & -\mu_{\mathrm{eff}}\\
-\mu_{\mathrm{eff}} & 0
\end{pmatrix}\ ,
\end{equation}
whilst the last block reads, in the $(\tilde{\phi}_{1}, \tilde{\varphi}_{2},
\tilde{\delta}_{1R}, {\tilde{\delta}_{2R}}, \tilde{S}, \B, \W_{L}^{0},
\W_{R}^{0})$ basis,
\begin{equation}
M_{\tchi^{0}}  = 
\begin{pmatrix}
0 & -\mu_{\mathrm{eff}} & 0 & 0 & -\mu_{d} & 0 & \frac{g_{L}v_{u}}{\sqrt{2}} & -\frac{g_{R}v_{u}}{\sqrt{2}} \\
- \mu_{\mathrm{eff}} & 0 & 0 & 0 & -\mu_{u} & 0 & -\frac{g_{L}v_{d}}{\sqrt{2}} & \frac{g_{R}v_{d}}{\sqrt{2}}\\
0 & 0 & 0 & \mu_{R} & \frac{\lambda_{R} v_{2R}}{\sqrt{2}} & g^\prime v_{1R} & 0 & -g_{R}v_{1R}\\
0 & 0 & \mu_{R} & 0 & \frac{\lambda_{R} v_{1R}}{\sqrt{2}} & -g^\prime {v}_{2R} & 0 & -g_{R}{v}_{2R}\\
-\mu_{d} & -\mu_{u} & \frac{\lambda_{R} {v}_{2R}}{\sqrt{2}} & \frac{\lambda_{R} v_{1R}}{\sqrt{2}} & \mu_S & 0 & 0 & 0\\
0 & 0 & g'v_{R} & -g^\prime{v}_{2R} & 0 & M_{1} & 0 & 0\\
\frac{g_{L}v_{u}}{\sqrt{2}} & -\frac{g_{L}v_{d}}{\sqrt{2}} & 0 & 0 & 0 & 0 & M_{2L} & 0\\
-\frac{g_{R}v_{u}}{\sqrt{2}} & \frac{g_{R}v_{d}}{\sqrt{2}} & -g_{R}v_{1R} & -g_{R}{v}_{2R} & 0 & 0 & 0 & M_{2R}
\end{pmatrix}\ ,
\end{equation}
where we have def\mbox{}ined $\mu_S=\lambda_S \frac{v_s}{\sqrt{2}}$, $\mu_{L,R}=\lambda_{L,R}\frac{v_{s}}{\sqrt{2}}$ and $\mu_{u,d}=\lambda_{3} \frac{v_{u,d}}{\sqrt{2}}\,$.

\section{Constraints on the spectrum and model parameters}
\label{sec:constraints}

We study in this work the collider signals associated with sneutrino dark matter LRSUSY scenarios at the LHC. As we shall argue in the fol\mbox{}lowing, the resonant production of RH sleptons via a $W_R$-boson exchange is a promising channel. Moreover, the decay chains of heavier superpartners to sneutrinos typically lead to multileptonic f\mbox{}inal states, and the corresponding SM background is small.

We have computed the particle spectrum with {\sc SPheno-3.3.8} \cite{Porod:2011nf}, the model f\mbox{}iles being generated with {\sc Sarah}~\cite{Staub:2013tta}. For
a reliable computation of the doubly-charged Higgs mass, we have used the dedicated method introduced in Ref.~\cite{Basso:2015pka}. We have then scanned the
parameter space to design our four benchmark scenarios. we describe in the next
subsections the constraints that we have imposed and the corresponding
phenomenological consequences of our benchmark design strategy.

\subsection{Right-handed gauge sector}
\label{sec:gaugesector}
Unlike in non-supersymmetric left-right symmetric extensions of the SM,
predictions for the masses of the right-handed gauge bosons exhibit an upper
limit, which depends on the SUSY breaking scale~\cite{Kuchimanchi:1995vk,Babu:2014vba}, as
the charge-conserving vacuum is not stable when $v_R\gg M_{\mathrm{SUSY}}$. However, such a
limit does  not hold if $B-L=0$ triplets stabilize the vacuum and the
right-handed gauge sector is extremely heavy, with masses of
$\mathcal{O}(10^{11})$~GeV~\cite{Aulakh:1997fq}.

Both the ATLAS and CMS collaborations have searched
for RH charged and neutral gauge bosons. Current bounds on such
additional gauge bosons are derived from both their hadronic and leptonic decay channels~\cite{Chatrchyan:2012gqa,CMS:2012zv,Khachatryan:2014dka,ATLAS:2015nsi,Sirunyan:2016iap} and are quite strong, the $W_R$ and $Z_R$ masses being constrained to lie above
about $2.7$~TeV. However, in the LRSUSY setup, the gauge bosons can easily
possess additional decay modes to a pair of lighter supersymmetric particles
(usually electroweakinos or sleptons) or to some of the new scalar bosons.
All these new modes invariably affect the total decay width and the branching ratios of these gauge
bosons so that the existing limits cannot be directly/blindly applied.

We define our benchmark points by setting the branching fraction of the $W_R$-boson to supersymmetric f\mbox{}inal states to be of $10$--$15\%$, while the total
branching ratio into SM final states is fixed to 65--70\%. This implies
that decays into Higgs states are as well possible. The current bound of $2.7$~TeV
obtained in the CMS dijet analysis~\cite{Sirunyan:2016iap} can thus be easily
relaxed, the choice $M_{W_R}=2.7$~TeV being perfectly viable.

We firstly adopt an optimistic benchmark scenario where the $W_R$-boson mass is
close to $2.7$~TeV, which gives, since the RH triplet VEV and the $Z_R$-boson mass are related,
\be
  M_{W_R}=2.7~{\rm TeV}\ , \qquad v_R = 5.7~{\rm TeV}
  \qquad\text{and}\qquad
  M_{Z_R} = 4.5~{\rm TeV} \ .
\ee
We secondly include in our study a more pessimistic benchmark point where the
$W_R$-boson mass is larger,
\be
  M_{W_R}=3.5~{\rm TeV}\ , \qquad v_R = 7.5~{\rm TeV}
  \qquad\text{and}\qquad
  M_{Z_R} = 5.9~{\rm TeV} \ .
\ee
The current LHC bounds on the existence of a $Z_R$ boson~\cite{ATLAS:2015nsi,Aaboud:2016cth,Khachatryan:2016zqb} are satisf\mbox{}ied in both cases.

\
Eventually, the LHC will probe higher $W_R$ masses. It has been shown that the
discovery of $W_R$ bosons with masses reaching up to $5$~TeV and their exclusion
for masses as large as 6~TeV could be achieved with $300$~fb$^{-1}$ of
proton-proton collisions at a centre-of-mass energy of
$\sqrt{s}=14$~TeV~\cite{Egede:1997dsa}. Our LRSUSY parameterization
allows for stable vacua featuring a heavy $W_{R}$ boson with a mass
ranging of up to about $8$~TeV. The total exclusion of $W_{R}$ bosons
predicted in LRSUSY models nevertheless requires a higher collision energy than
the one available at the LHC, as there will always remain parts of the parameter
space where the $W_R$ boson can
escape detection (see also discussion in Sec.~\ref{subsec:dch}).

\subsection{The Higgs sector}
\label{subsec:dch}

In LRSUSY, the doubly-charged Higgs sector plays a central role, not only in terms of 
the construction of the model, but also for its phenomenology.
The doubly-charged Higgs mass matrix has a negative eigenvalue at tree-level
once the neutral component of the triplet gets a VEV, and one-loop corrections
must be included for stabilizing the scalar potential~\cite{Babu:2008ep}. The original work
relied on the lepton-slepton contributions and hence the couplings
of the leptons to the RH triplet Higgs superfield must be taken large for at least
one generation. The same couplings however govern the decays of the
doubly-charged Higgs boson and it is important to verify the consistency with
the various LHC bounds. The latter are strong, with the exception of the case
in which the doubly-charged Higgs boson decays into a ditau
final-state~\cite{Chatrchyan:2012ya,CMS:2016cpz,CMS:hig16036}. Such searches
have so far managed to push strong bounds on setups where the lightest
doubly-charged Higgs boson is of a LH triplet nature. These bounds can be
evaded in typical LRSUSY scenarios for RH doubly-charged Higgs bosons, since the associated
production of $\delta^{\pm\pm}\delta^{\mp}$ through $W_L^{\pm}$ is not possible and in the pair production the triplet Higgs
couples to the $Z$-boson only through the $B-L$ and $W_{3R}$ components leading to
a suppression in the production cross section compared to the left-handed triplets.

Whereas one may assume that the
triplet Higgs superfield mostly couples to third generation (s)leptons, non-zero
couplings to the other generations are needed to generate masses for the
RH neutrinos. In order to evade all doubly-charged Higgs LHC constraints, we
fix the model free parameters
in a way in which the branching ratio of the doubly-charged Higgs 
boson into muons and electrons stays below $10\%$.

Further constraints arise from the sign of the overall one-loop correction to the
doubly-charged Higgs mass, which depends on the slepton masses and $v_R$. If the
slepton masses are much smaller than $v_{R}$, the correction will be
negative and worsen the problem of the negative mass
eigenvalue~\cite{Babu:2014vba}. The large value of $v_{R}$ that we have
adopted hence disfavors a light slepton option. It has however been recently found that the
gauge and Higgs sectors can also signif\mbox{}icantly contribute to the
doubly-charged Higgs mass, which opens up a window for lighter RH
sleptons and sneutrinos~\cite{Basso:2015pka}.

To obtain a heavy enough doubly-charged Higgs boson, we set the $\lambda_{R}$
parameter to a large value, which leads to a large contribution to the doubly-charged
Higgs-boson mass from the Higgs sector. We moreover make the electroweakinos
rather heavy for benchmarks featuring a sneutrino LSP in order to avoid a
neutralino LSP.
As mentioned in Sec.~\ref{sec:lrsusy}, we assume that the LH Higgs triplets are
inert, so that the corresponding masses are determined by the soft
supersymmetry-breaking parameters. Being less relevant for the phenomenology of
interest, we set their value larger than $1$~TeV.

Whereas in the work of Ref.~\cite{Basso:2015pka}, the
highest values for $m_{\Delta^{\pm\pm}}$ turned out to be around $650$~GeV,
stable vacua can still be achieved with doubly-charged Higgs boson masses
slightly above $800$~GeV. The projected LHC sensitivity for 100~fb$^{-1}$ of
luminosity shows a reach for a potential $3\sigma$ discovery that extends up to
$950$~GeV when the doubly-charged Higgs boson exclusively decays into electrons
or muons~\cite{Babu:2016rcr}.
The doubly-charged Higgs boson limits stemming from decay modes with tau leptons
are not as stringent, due to the efficiency of tau identification. Excluding a
$800$~GeV doubly-charged Higgs boson decaying solely to same-sign taus would
require an improvement of two orders of magnitude with respect to the latest CMS
bound~\cite{CMS:hig16036}, which will be challenging even with $3000$~fb$^{-1}$.
It is therefore uncertain whether the LHC will be able to exclude the model
on the basis of doubly-charged Higgs boson searches only due to the structure
of the Yukawa couplings (and the various doubly-charged Higgs boson branching
ratios). Moreover, if the
vacuum is stabilized by the introduction of $B-L=0$ triplets, the doubly-charged
Higgs boson can be heavier and outside the reach of the LHC. Furthermore,
the discovery of a doubly-charged Higgs boson would not be a signal specific to
LRSUSY setups and should be used in conjunction with other measurements to draw
conclusive LRSUSY statements.
The discovery of a doubly-charged scalar field along with doubly-charged 
higgsinos and a RH gauge boson would be a strong hint towards establishing a
left-right supersymmetry without discovering any other SUSY particle, simply by 
virtue of the robustness of the signal. Such signals have therefore already been studied widely in the literature~\cite{Dutta:1998bn, Chacko:1997cm, Demir:2008wt, Demir:2009nq, Babu:2013ega, Alloul:2013raa}, and in this work we focus on another sector of the model.

The upper limit on the tree-level mass of the SM-like Higgs-boson can be much larger than in the MSSM by the virtue of the 
extended gauge sector. If $g_{L}=g_{R}$, one finds~\cite{Babu:1987kp}
\be
m_h^{\rm tree} \leq \sqrt{2}m_{W}\simeq 113.7{\rm~GeV}\ ,
\ee a value that can
be easily lifted to about $125$~GeV by incorporating the radiative corrections
and by adjusting the stop masses and mixing. The latter depends on
$\tan \beta$, and for values close to one, the tree-level mass of the lightest
scalar boson tends to vanish, like in the MSSM.
However, the LRSUSY $D$-terms can increase the tree-level Higgs mass beyond values
that are typical from the MSSM.
Focusing on the rest of the Higgs sector, the second $CP$-even, the lightest
$CP$-odd and the lightest singly-charged Higgs boson can have masses below or slightly
above 1~TeV. Close to the alignment limit, their dominant decay modes involve third
generation fermions and the related LHC reach is thus similar as for the heavier
states of the MSSM.

We choose a moderate value for  $\tan \beta$ so that the bounds stemming
from both the direct heavy Higgs-boson searches in the $H/A\rightarrow \tau\tau$
channel~\cite{Khachatryan:2014wca,Aaboud:2016cre} are weaker and the
contributions to the rare $B_{s}\rightarrow \mu\mu$ decay are smaller than for large $\tan \beta$. 
This has an additional advantage to suppress the mixing in the neutral Higgs sector,
which may challenge the SM-nature of the lightest state and lead to
a large deviation from the SM for the $h\rightarrow b\bar{b}$ branching ratio~\cite{Frank:2014kma}.

Turning to singly-charged Higgs bosons, indirect constraints originating
from $b\rightarrow s\gamma$ data~\cite{Saito:2014das} suggest that they
must be heavy~\cite{Hermann:2012fc}, at least if there are
no cancellations in the SUSY loop-contributions to the single-charged Higgs-boson mass. This can be
accommodated in LRSUSY setups if $\tan \beta_{R}$ deviates from one and if
$v_R$ is large. Such a deviation will subsequently impact one of the diagonal
elements of the doubly-charged Higgs mass matrix, making it smaller, and render
the task of satisfying the doubly-charged Higgs mass constraints more
difficult.
We therefore adopt
\be
\tan \beta_{R}\simeq 1.05\ ,
\ee
which, with our chosen values for $v_{R}$, pushes all the
MSSM-like Higgs states to be heavier than current LHC bounds. They have masses squared proportional to $g_R^2v_R^2(\tan^2\beta_R-1)$ and are hence at most just above the TeV scale. 
Moreover, all
additional scalar bosons  
have masses of the order
of $v_R$, $v_S$, or of the LH triplet soft mass parameters and  hence are a lot heavier.

\subsection{The neutrino sector}
\label{subsec:RHneutr}
The RH $W_{R}$-boson directly decays into RH leptons and neutrinos, provided
this decay channel is open. In this case, RH neutrinos could be significantly
produced via the resonant production of a $W_{R}$-boson, which offers a handle
to constrain
the masses of the RH neutrinos as a function of the $W_{R}$-boson mass. As the
RH neutrino subsequently decays through the $N_{R}\to \ell W_{R}^{*}\to \ell jj$
channel, the corresponding collider signal ($p p \to N_R \ell \to \ell\ell j j$)
is made of two charged leptons and
two jets, the Majorana nature of the neutrino implying a similar amount of
same-sign dilepton and opposite-sign dilepton events~\cite{Keung:1983uu}. CMS
has relied on these considerations to derive simultaneously bounds on the RH
neutrinos and gauge bosons and express them as contours in the
$(M_{W_{R}}, M_{N_{R}})$ mass plane, both for the electronic and muonic
channels~\cite{Khachatryan:2014dka}. The bounds are stronger when the neutrino
masses lie in the [$400$~GeV, $1$~TeV] mass window, the $W_R$-boson being
constrained to be heavier than 3~TeV. In contrast, for setups with neutrinos
lighter than $200$~GeV, like in the benchmark scenarios used in this study, the
$W_{R}$-bosons are instead only constrained by resonance searches in the dijet mode
(see Sec.~\ref{sec:gaugesector}). We additionally verify that the (weaker) lower
bounds extracted from LEP data are fulfilled, which requires the RH Majorana
neutrino masses to be above about 90~GeV~\cite{Achard:2001qw}.

As indicated in Sec.~\ref{subsec:dch}, the doubly-charged Higgs-boson is
enforced to decay into an electron or a muon pair with a small branching ratio.
This simultaneously drives the RH neutrino masses to low values, as they arise
mostly from the Higgs triplet couplings. The lepton-slepton contributions to the
doubly-charged Higgs mass cannot however be too negative to ensure that the LHC
direct search limits are satisfied, which consequently
provides an upper limit on these couplings. We hence set the Yukawa coupling matrix
$Y_{L}^{4}$ to be diagonal, and include a hierarchy on the diagonal entries so
that the doubly-charged Higgs-boson ditau decay mode is associated with a
branching ratio larger than $90\%$.

\subsection{The neutralino and chargino sector}
\label{subsec:lim-charg-neutr}
We investigate in this work scenarios in which a sneutrino is a DM candidate,
so that neutralinos and charginos must be heavier. The mass of the 
gaugino-dominated states can be made heavy by setting the corresponding
soft masses to large values and we use the singlet superpotential self-coupling
$\lambda_{S}$ to prevent the singlino-dominated state to be too light and thus
the LSP. The higgsino states are in contrast automatically heavy by
virtue of the large $v_{1R}$, $v_{2R}$ and $v_S$ values.

The LHC phenomenology connected to LRSUSY neutralino and chargino states has
been recently analysed in Ref.~\cite{Alloul:2013fra}, where it has been shown
that the leptonic channels are the best probes for LRSUSY neutralinos and
charginos. The production rates are in general larger than in the MSSM
for not too heavy gauginos, so that this additionally offers handles to
distinguish the LRSUSY case from the MSSM. For a comparative study with cases where the neutralino is the LSP, we focus on
LRSUSY realizations where the lightest neutralino is bino-dominated. In this
case, we fix the bino soft-mass $M_{1}$ to a value yielding a DM relic
density as measured by the Planck satellite.

\subsection{Benchmark point definitions}
\label{sec:bp}
As mentioned in Sec.~\ref{sec:gaugesector}, we consider two sneutrino LSP
benchmark points, the first one (BP1) featuring a lighter $W_R$-boson with
$M_{W_R}\simeq 2.7$~TeV, and a second one (BP2) featuring a heavier RH gauge
boson with $M_{W_R}\simeq 3.5$~TeV.
We additionally define two comparative scenarios BP3 and BP4 where the lightest
neutralino is bino-like and the LSP, for the same $W_R$-boson masses of 2.7 and
3.5~TeV respectively.

\begin{table}
\setlength\tabcolsep{2pt}
\begin{tabular}{|c c|c c|}
\hline
Parameter & Value & Parameter & Value\\
\hline
$\lambda_{L}$ & $0.4$ & $\lambda_{R}$ & $0.9$\\
$\lambda_{S}$ & $-0.5$ & $T_{R}$ & $-2$~TeV\\
$T_{S}$ & $-2$~TeV &  $T_{3}$ & $1$~TeV\\
$M_{{\Delta 1L},{\Delta 2L}}^{2}$ & $2$~TeV$^{2}$ & $M_{3}$ & $3.5$~TeV \\
$(Y_{L}^{4})_{ii}$ & ($0.019,0.022,0.1$) & $\xi_{F}$ & $-5000$~GeV$^{2}$ \\
\hline
\end{tabular} \hspace{0.05cm}\begin{tabular}{|c|c c c c|}
\hline
Parameter & BP1 & BP2 & BP3 & BP4\\
\hline
$\tan \beta$ & $6.5$ & $8$ & $7$ & $7$\\
$\tan \beta_R$ & $1.05$ & $1.05$ & $1.04$ & $1.04$\\
$v_R$ (TeV) & $5.7$ & $7.5$ & $5.7$ & $7.5$\\
$v_S$ (TeV) & $7$ & $10$ & $7$ & $8$ \\
$\lambda_{3}$ & $0.15$ & $0.10$ & $0.10$ & $0.08$ \\
$M_{2L,R}$ (GeV) & $1200$ & $900$ & $700$ & $700$ \\
\hline
\end{tabular}
\caption{\sl Benchmark scenario definitions. Parameter values common to all
  benchmark points are shown in the left panel, while benchmark-specific choices
  are shown in the right panel. We moreover set
  $\lambda_4=\lambda_5=T_{L}=T_{4}=T_{5}=0$ for simplicity.}
\label{tb:fixpar}\end{table}

We present the values of the most important model parameters in
Table~\ref{tb:fixpar}, as extracted from our scanning procedure, and the
relevant part of the particle spectrum in
Table~\ref{tb:spectrum}. The latter also includes the values of several
low-energy observables that have been used to constrain the model, enforcing the
predictions to agree within two standard deviations with the current experimental values. Whereas we only list the
masses for the three lightest neutralino states, the next two
lightest neutralinos are nearly degenerate in mass with the
$\tilde{\chi}^{0}_{3}$ neutralino. In
the BP1 case, the four lightest states are all higgsino-dominated and the
f\mbox{}if\mbox{}th one is bino-dominated, whereas for the other benchmark
points, the bino-dominated state is the lightest and the next four neutralinos
are higgsino-dominated.

\begin{table}
\begin{center}
\setlength\tabcolsep{5pt}
\begin{tabular}{|c|c c c c|}
\hline
Particle & BP1 & BP2 & BP3 & BP4\\
\hline
$h$ & $125.2$ & $125.5$ & $124.8$ & $125.3$\\
$H_2$ & $551.1$ & $748.5$ & $492.4$ & $657.9$\\
$H_3$ & $1958$ & $2076$ & $1949$ & $2363$ \\
$A_1$ & $551.1$ & $748.5$ & $492.4$ & $657.9$ \\
$H^{\pm}_{1}$ & $563.7$ & $757.7$ & $506.0$ & $668.1$\\
$H^{\pm\pm}_{1}$ & $339.1$ & $494.6$ & $431.7$ & $509.8$\\
$W_{R}^{\pm}$ & $2668$ & $3510$ & $2668$ & $3510$\\
$Z'$ & $4476$ & $5889$ & $4476$ & $5889$\\
\hline
$\nu_{Re}$ & $104.2$ & $136.8$ & $104.7$ & $137.6$ \\
$\nu_{R\mu}$ & $120.7$ & $158.4$ & $121.2$ & $159.2$\\
$\nu_{R\tau}$ & $548.5$ & $719.6$ & $550.8$ & $724.1$ \\
$\tilde{\nu}_{I\tau}$ & $266.5$ & $271.6$ & $416.0$ & $299.7$\\
$\tilde{\nu}_{Ie}$  & $813.8$ & $663.6$ & $632.2$ & $896.3$\\
$\tilde{\nu}_{I\mu}$ & $856.9$ & $716.2$ & $792.0$ & $947.3$\\
$\tilde{\nu}_{Re}$ & $1301$ & $1454$ & $1159$ & $1488$ \\
$\tilde{\nu}_{R\mu}$ & $1331$ & $1566$ & $1312$ & $1590$ \\
$\tilde{\nu}_{R\tau}$ & $2262$ & $2983$ & $2269$ & $2742$ \\
$\tilde{e}_{R}$ & $931.7$ & $813.8$ & $773.3$ & $1011$ \\
$\tilde{\mu}_{R}$ & $931.7$ & $928.2$ & $947.3$ & $1105$ \\
$\tilde{\tau}_{R}$ & $1399$ & $1837$ & $1449$ & $1678$ \\
\hline
$\tilde{\chi}^{0}_{1}$ & $731.1$ & $609.8$ & $61.9$ & $62.4$ \\
$\tilde{\chi}^{0}_{2}$ & $750.6$ & $711.3$ & $486.6$ & $447.2$ \\
$\tilde{\chi}^{0}_{3}$ & $750.9$ & $716.3$ & $501.1$ & $459.3$ \\
$\tilde{\chi}^{\pm}_{1}$ & $744.0$ & $703.7$ & $487.5$ & $447.8$ \\
\hline
BR($b\rightarrow s\gamma$) & $3.04\times 10^{-4}$ & $3.10\times 10^{-4}$ & $3.03\times 10^{-4}$ & $3.08\times 10^{-4}$ \\
BR($B_{s}\rightarrow \mu\mu$) & $2.74\times 10^{-9}$ & $3.68\times 10^{-9}$ & $3.44\times 10^{-9}$ & $2.71\times 10^{-9}$ \\
$\Delta a_{\mu}$ & $1.2\times 10^{-10}$ & $1.5\times 10^{-10}$ & $2.1\times 10^{-10}$ & $1.9\times 10^{-10}$ \\
\hline
\end{tabular}
\end{center}
\caption{\sl The relevant particle spectrum of the four adopted benchmark points, presented
together with the values of several low-energy observables suitable for
constraining the model. All of the sleptons given here are RH. The subscripts $R$ and $I$ in sneutrinos refer to real and imaginary parts, respectively. All masses are given in GeV.}
 \label{tb:spectrum}
\end{table}

\section{Dark matter phenomenology}
\label{sec:dm}
In this section, we respectively focus on scenarios where the LSP is a sneutrino
(Sec.~\ref{subsec:sneutrinodm}) and a neutralino
(Sec.~\ref{subsec:neutralinodm}). We explore the impact of small deviations from
the benchmarks introduced in the previous section on the dark matter relic
density and investigate how well this agrees with the observed
value~\cite{Ade:2015xua},
\be
  \Omega_{\rm DM}h^2=0.1198\pm 0.0015\ ,
\ee
$\Omega_{\rm DM}$
being as usual the dark matter energy density evaluated relatively to the
critical energy density of the universe and $h$ the reduced Hubble parameter.
We additionally investigate the robustness of the direct DM detection bounds as
a function of the model parameters.

\subsection{Sneutrino dark matter}
\label{subsec:sneutrinodm}

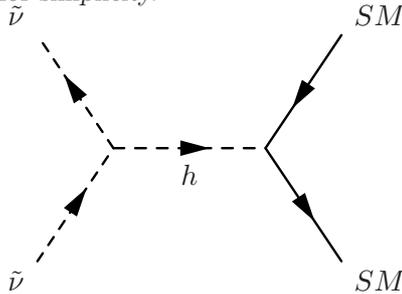
\begin{figure}
\begin{center}
\unitlength=1mm
\begin{fmffile}{snu1}
\begin{fmfgraph*}(50,30)
\fmfleft{i1,i2}
\fmfright{o1,o2}
\fmflabel{$\tilde{\nu}$}{i1}
\fmflabel{$\tilde{\nu}$}{i2}
\fmflabel{$SM$}{o1}
\fmflabel{$SM$}{o2}
\fmf{scalar}{i1,v1,i2}
\fmf{fermion}{o2,v2,o1}
\fmf{scalar,label=$h$}{v1,v2}
\end{fmfgraph*}
\end{fmffile}
\end{center}
\caption{\sl Dominant DM annihilation process for scenarios where the sneutrino
   is the LSP. The final state generically denotes any SM particle the Higgs
   boson couples to.\label{fig:snuann}}
\end{figure}

If the particle spectrum is such that there is no DM co-annihilation channels
significantly relevant, sneutrino LSP mostly annihilates via a SM-like
Higgs-boson exchange in the $s$-channel, as depicted in
F\mbox{}ig.~\ref{fig:snuann}. For setups where RH neutrinos are lighter than
sneutrinos, a $t$-channel neutralino exchange diagram also exists, although it
turns out to be suppressed for heavy neutralinos. We concentrate in this
work on scenarios where the RH neutrinos are always heavier than the LSP.

Dark matter relic density and direct detection constraints have been calculated
with  {\sc MadDM v2.0}~\cite{Backovic:2015cra}, and we have used
{\sc MicrOmegas}~\cite{Belanger:2014vza} to validate our findings. Due to the
dominance of the relic density on the $s$-channel diagrams given in
Fig.~\ref{fig:snuann}, it only depends on the RH sneutrino mass. Existing
bounds are found to be satisf\mbox{}ied with sneutrinos having a mass lying in
the [250~GeV, 290~GeV] range (provided all co-annihilation channels are
negligible). We adjusted the lightest sneutrino mass to lie in this range for our 
benchmarks. For the exact BP1 and BP2 parameters, we obtain a relic density
prediction of $\Omega_{\mathrm{DM}}h^{2}=0.119$ and $0.116$ for the BP1 and BP2
point respectively. In addition, the spin-independent cross sections for
DM-nucleon scattering is smaller than $2.5\times 10^{-10}$~pb for both
benchmark scenarios, which agrees with the current bounds.

If the sensitivity of direct detection experiments increases by a factor of
three, we should either see a signal or exclude typical benchmarks like BP1 or BP2. As the
DM-nucleon scattering is mediated mostly via the SM-like Higgs boson, a direct
detection signal will be largely unaffected by the details of the unknown
particle spectrum and thus provides a robust way of testing right sneutrino DM
in LRSUSY models. In presence of coannihilations or large mixing
between the left and right sneutrino sectors, the observed relic density may point
to another mass range than in our case. Moreover, if heavier sneutrinos are allowed, the
direct detection bounds are weaker and the exclusion will be more difficult.

We observe a remarkable feature originating from the left-right symmetry. By
construction, the RH sneutrino fields are a part of $SU(2)_R$ doublets while the
SM-like Higgs boson is in contrast originating mainly from the Higgs bidoublets.
As the RH neutrino and Higgs bidoublet fields couple very strongly, the
annihilation cross section is large enough for ensuring a correct relic
density, even if the associated process occurs non-resonantly. The RH sneutrino
coupling to the $Z$-boson is on the other hand weak enough for preventing dark
matter annihilation to be too ef\mbox{}f\mbox{}icient. As a consequence,
the only relevant parameters driving the relic density are the $SU(2)_R$ gauge
coupling (see Eq.~\eqref{eq:snucoupling}) and the LSP mass, so that a correct
relic density can be obtained for a broad range of LRSUSY realizations. This
drastically differs from cases where RH sneutrinos are gauge-singlet and where a
resonant annihilation is needed to guarantee a correct relic density, like in
the next-to-minimal supersymmetric standard model (NMSSM) extended with RH
neutrinos~\cite{Cerdeno:2008ep,Cerdeno:2009dv,Chatterjee:2014bva,%
Cerdeno:2014cda}, or when the dark matter candidate is a mixture of LH and RH
sneutrinos~\cite{ArkaniHamed:2000bq,Arina:2007tm,Belanger:2010cd,Arina:2015uea}.
In both these cases, extra free parameters are available to tune the relic
density to match the experimental bounds, in contrast with the LRSUSY setup
unless one includes a large left-right mixing in the sneutrino sector.

The main DM annihilation channel proceeding via a Higgs-boson exchange, this
consequently implies that the main sneutrino pair-production mode at the LHC
will proceed via an $s$-channel Higgs-boson exchange as well. Typical DM
searches relying on initial-state radiation would then be insensitive to this
LRSUSY setup, the order of magnitude of the corresponding cross section being at
the attobarn level due to a suppression by the weakness of the Higgs interactions with
the QCD partons and the non-resonant configuration driven by the Higgs-boson and
RH sneutrino mass difference. RH sneutrino are hence dominantly produced
from the decay of other particles. One obvious candidate is the heavier
MSSM-like Higgs state, but its (gauge) couplings to the sneutrinos whose form is similar to
Eq~\eqref{eq:snucoupling} vanishes in the alignment limit. The dominant LHC
RH sneutrino production mode therefore proceeds via the resonant production of a $W_R$-boson.

\subsection{Neutralino dark matter}
\label{subsec:neutralinodm}

LRSUSY neutralino dark matter has been already discussed in the past, but under
assumptions different from ours. Ref.~\cite{Demir:2006ef} considers
neutralinos that are pure gauge eigenstates, so that their results must be
generalized to the case where neutralinos are admixtures of electroweakinos and
higgsinos, whereas Ref.~\cite{Esteves:2011gk} has built a LRSUSY model where
$B-L=0$ Higgs triplets are included and RH Higgs triplets VEVs are decoupled,
forbidding the $W_R$-induced production of pairs of superpartners.

In our LRSUSY parameterization, the composition of the lightest neutralino
depends on the soft gaugino masses. We have chosen $M_{1}$ to be the smallest
gaugino mass parameter to guarantee a bino-dominated LSP, and fix its value in
order to obtain a relic density prediction in agreement with data. The $M_1$
parameter is nonetheless connected to the $U(1)_{B-L}$ gaugino (that we
abusively call bino), so that it does not couple to the light gauge bosons and
the bidoublet Higgs fields. The bino however mixes with the other gauginos,
which
ensures non-vanishing couplings to the $Z$-boson and the SM-like Higgs-boson.

The resulting relic density is in general too large and we need a resonant
contribution to increase the DM annihilation cross section. For this reason, our
neutralino LSP benchmarks feature a $\tilde\chi_1^0$ mass slightly below half
the Higgs-boson mass $m_h/2$. The kinematically allowed $h\to\tchi_1^0\tchi_1^0$
decay is suppressed since the LSP is bino-dominanted and the bidoublets are not
charged under $U(1)_{B-L}$. The corresponding branching ratios for the BP3 and
BP4 cases are found to be about $4\times 10^{-4}$, which is of the same order as
the SM Higgs-boson invisible branching ratio, and the associated relic density
is respectively $\Omega_{\mathrm{DM}}h^{2}=0.107$ and $0.124$ for the two
benchmark scenarios. Furthermore,
the spin-independent and spin-dependent nucleon-DM scattering cross sections
are of $3\times 10^{-11}$~pb and $2\times 10^{-6}$~pb for both benchmark points,
which satisfies current direct detection bounds~\cite{Akerib:2016vxi}.

\section{Collider phenomenology at the LHC}
\label{sec:signals}

\subsection{Analysis strategy for discovering LRSUSY at the LHC}

In the MSSM, the production of weakly interacting superpartners in
proton-proton collisions at a center-of-mass energy of 13~TeV is limited, so
that current search limits for sleptons and electroweakinos are weaker than for
the strongly interacting sector. These searches additionally rely on a very high
LHC luminosity to be sensitive to superparticles lying in the 1~TeV mass regime.
Moreover, production cross sections for RH scalar partners are smaller than for
LH partners for a given superparticle mass. This is one of the most crucial
differences for the LRSUSY case, RH scalar production cross sections being here
enhanced thanks to the gauging of the RH sector, on top of the fact that a RH
sneutrino can be a good DM candidate.

Right-handed slepton and sneutrino production at the LHC is mediated by heavy
RH gauge boson exchanges. The corresponding rates are
enhanced if resonant configurations are reached, so that the LHC is possibly
sensitive to high mass scales. In this section, we make use of resonant
slepton and sneutrino production to show how robust and clean these signals can
be and how they can provide handles for pushing the LHC reach for the
weakly-interacting sector beyond 1~TeV. We consider the production process
\begin{equation}
  p \, p \rightarrow \sum \tilde{\nu} \ \tilde{\ell}
\end{equation}
where we sum over all possible final states, \ie, we include three generations
of sleptons and of scalar and pseudoscalar RH sneutrinos. We observe that the bulk of the cross
section originates from on-shell $W_R^{\pm}$ production followed by its decays
into a slepton-sneutrino f\mbox{}inal state. Our predictions rely on the UFO
libraries~\cite{Degrande:2011ua} outputted by {\sc Sarah} to generate the
relevant hard-scattering matrix elements with
{\sc MadGraph5}\_aMC@NLO~\cite{Alwall:2014hca}. More precisely, we convolute 
these matrix elements with the NNPDF 2.3 sets of parton
densities~\cite{Ball:2014uwa} to obtain the leading-order cross section
values indicated in Table \ref{tb:branching}, the branching ratios being those
returned by {\sc SPheno}. We find that the smaller $W_R$-boson production
cross sections for the BP2 and BP4 cases are partly compensated by the larger
branching ratios, but will also feature a sleptonic decay phase space
configuration where a slightly harder transverse-momentum ($p_T$) is
expected for the heavy gauge boson decay products.

\begin{table}
\begin{center}
\renewcommand{\arraystretch}{1.4}
\setlength\tabcolsep{7pt}
\begin{tabular}{|l|c c c c|}
\hline
 & BP1 & BP2 & BP3 & BP4\\
\hline
$\sigma(pp\rightarrow W_R)$ (fb) & $245$ & $38$ & $245$ & $38$ \\
BR($W_R \rightarrow \tilde{\nu}_{I\tau} \tilde{\ell}_{\tau}$) & $0.52\%$ & $0.52\%$ & $0.38\%$ & $0.61\%$ \\
BR($W_R \rightarrow \tilde{\nu}_{Ie} \tilde{\ell}_{e}$) & $0.64\%$ & $1.06\%$ & $0.80\%$ & $0.82\%$ \\
BR($W_R \rightarrow \tilde{\nu}_{I\mu} \tilde{\ell}_{\mu}$) & $0.60\%$ & $0.98\%$ & $0.57\%$ & $0.74\%$ \\
BR($W_R \rightarrow \tilde{\nu}_{Re} \tilde{\ell}_{e}$) & $0.21\%$ & $0.60\%$ & $0.42\%$ & $0.47\%$ \\
BR($W_R \rightarrow \tilde{\nu}_{R\mu} \tilde{\ell}_{\mu}$) & $0.24\%$ & $0.47\%$ & $0.19\%$ & $0.36\%$ \\
\hline
$\sigma\times \sum$BR($W_{R}\rightarrow \tilde{\nu}\tilde{\ell}$) (fb) & $5.4$ & $1.4$ & $5.8$ & $1.1$ \\
\hline
\end{tabular}
\end{center}
\caption{\sl $W_R$-boson production cross sections for proton-proton collisions
  at a center-of-mass energy $\sqrt{s}=13$~TeV, and $W_R$ branching ratios to
  sleptonic f\mbox{}inal states.\label{tb:branching}}
\end{table}

As $\mathcal{O}(100$~fb$^{-1})$ of integrated luminosity can be recorded by
the LHC current and future runs, our cross section results show that a fairly
reasonable number of signal events could be expected for our typical benchmark
scenarios. Slepton production could hence be a promising mode to look for
LRSUSY signals provided the SM background could be reduced.

We start with processes in which the lightest sneutrino is produced.
For the BP2, BP3 and BP4 scenarios, charged sleptons almost always decay into a
charged lepton of the same flavor and a light neutralino $\tchi^{0}_{1}$.
In contrast, in the BP1 scenario, this decay mode only occurs with a probability of about 30\% and
sleptons mostly decay into a $\ell \tilde{\chi}^{0}_{5}$ final state with a
branching ratio of 70\%. The heavier $\tilde{\chi}^{0}_{5}$ neutralino then
decays entirely into a final-state system made of a $W$-boson and the lightest
chargino that further decays, with a 100\% probability, into the LSP and a tau
lepton. When the lightest neutralino $\tilde{\chi}^{0}_{1}$ is the LSP for the
BP3 and BP4 scenarios, it decays invisibly for the BP2 scenario so that the
three benchmarks will feature a similar signature,
\be
  {\rm BP2-BP3-BP4}: \qquad
  p p \to  \ W_R \ \to \ \sum\tilde \ell\ \tilde\nu \ \to\
    1\ell + \Emiss \ .
\label{eq:bp234dec}\ee
For the case of the first benchmark scenario, extra signatures have to be
considered, in particular as a same-sign dilepton signal could arise due to the
Majorana nature of the intermediate $\tchi^{0}_{5}$ state,
\be
  {\rm BP1}: \qquad
  p p \to  \ W_R \ \to \ \sum\tilde \ell\ \tilde\nu \ \to\
    \ell + \Emiss 
    \quad\text{or}\quad
    \ell + \tau + W + \Emiss\ .
\ee
The golden same-sign dilepton signal however suffers from the low $W$-boson
leptonic branching fraction. On the other hand, heavier sneutrinos produced in
association with charged sleptons decay down to a chargino and a charged lepton
($e$ or $\mu$) almost half of the time, which suggests that at least one highly
energetic final-state charged electron or muon can be expected in such
processes.

\begin{figure}
\includegraphics[width=3.0in]{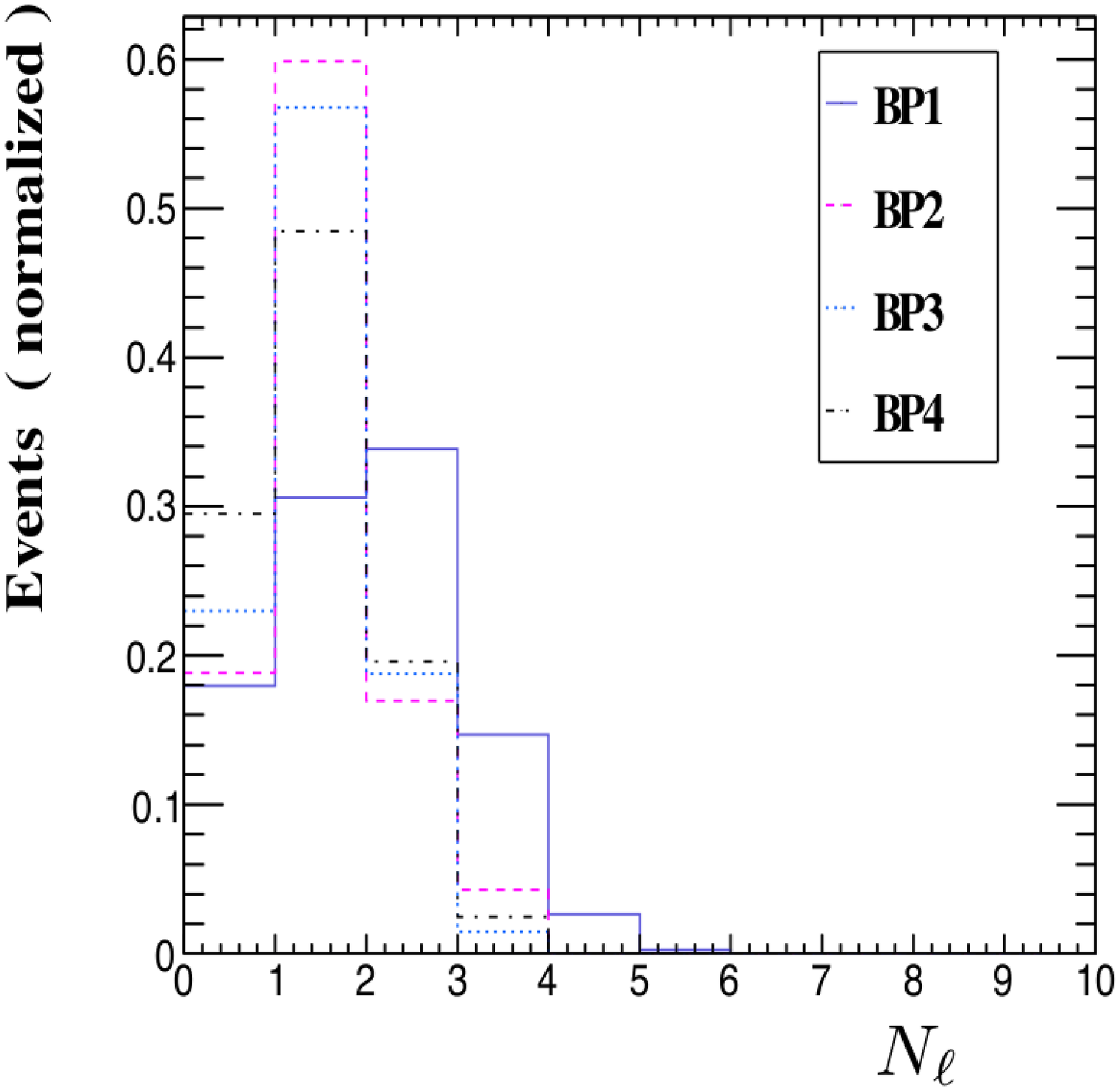}
\includegraphics[width=3.0in]{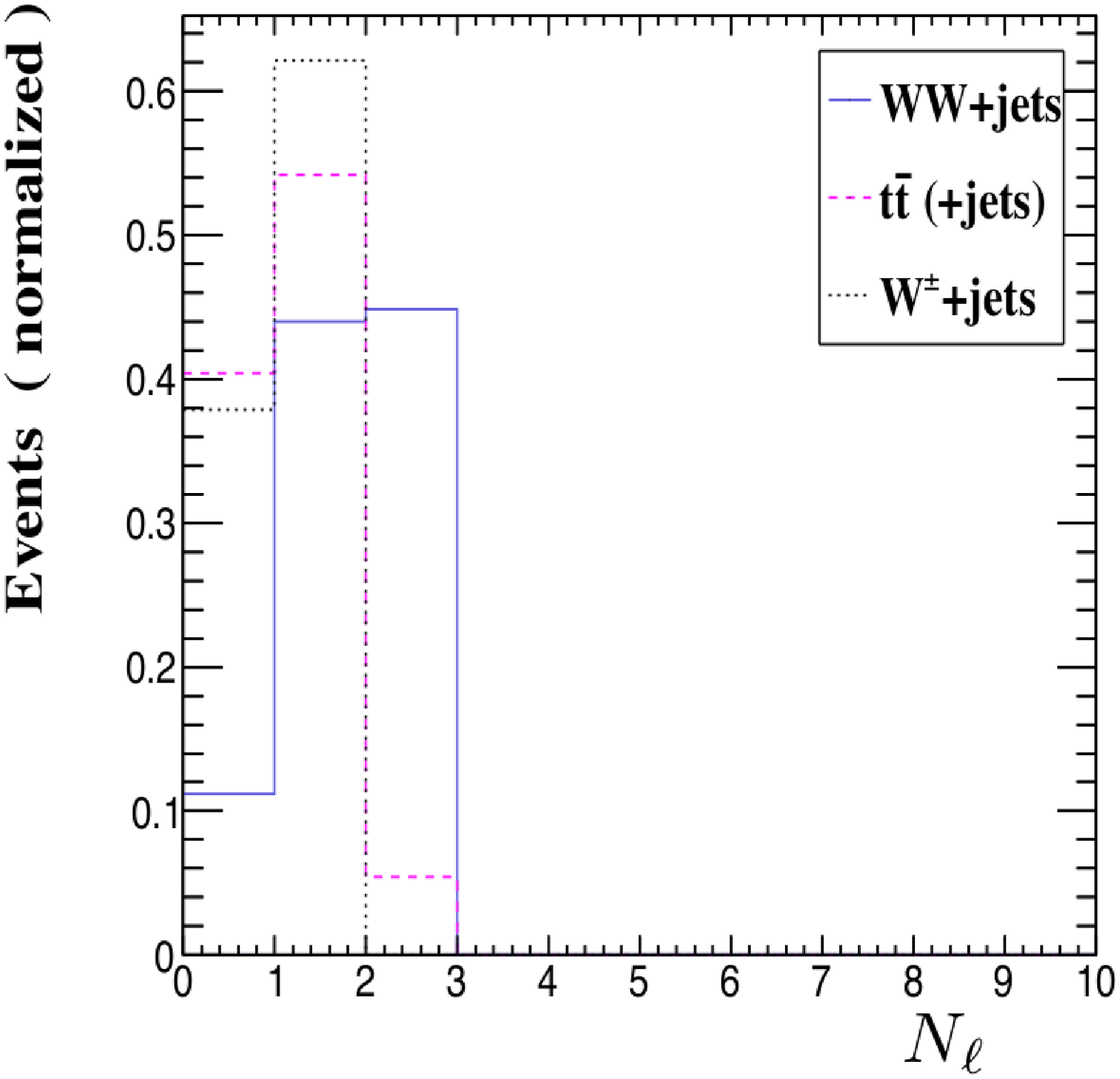}
\caption{\sl Parton-level charged lepton multiplicity distributions for the LRSUSY signals
   (left) and for the dominant SM backgrounds (right), all curves being
   normalised to 1. }
\label{fig:nlepmult}
\end{figure}

For simplicity, we ignore all electronic or muonic tau decays in the rest of our
analysis, although $\tau$-enriched final states would be more frequently
produced as the sneutrino LSP is of the $\tau$ flavor.  In
Fig.~\ref{fig:nlepmult} we present 
the  charged lepton multiplicity of the signal that is expected for the different
benchmark points (left) as well as for the dominant source SM background
(right). The predictions are calculated at the parton level and the results have
been normalized to 1. This suggests to consider two possible signal signatures,
\be\bsp
 (i) &\quad \geq 1\ell + n j + \Emiss \qquad\text{with}\quad   n \leq 3 \ , \\
 (ii)&\quad \geq 2\ell + n j + \Emiss \qquad\text{with}\quad   n \leq 3 \ ,
\esp\label{final_states} \ee
where $\ell = e^\pm$ or $\mu^\pm$ and the constraint on the jet multiplicity
originates from the topology of the signal that is poor in final-state jets.
Even if appealing, the second LRSUSY signal handle~($ii$) may be challenging to
use, the associated production rate being expected to suffer from an important
suppression relative to the first signature~($i$). We therefore focus, for
this pioneering study, on final-state systems containing one or more charged
leptons accompanied by a large amount of missing transverse momentum $\Emiss$
and a small number of jets.
The dominant source of SM background is expected to consist of
charged-current Drell-Yan-like production (in association with jets) where typical
final-states feature a single hard lepton (as shown in the right panel of
Fig.~\ref{fig:nlepmult}) and missing energy carried by the final-state neutrino.
Top quark-antiquark pair production also contributes when leptonic top quark
decays are considered, as well as various diboson production channels that
traditionally give rise to lepton-enriched final-states featuring missing energy
as well. Neutral current Drell-Yan events could in principle
contribute, but the corresponding final-state does not usually exhibit a large
amount of missing energy so that it can be rejected quite strongly with an
appropriate missing energy selection, such as the one performed in our analysis
(see below). We have also verified that triboson contributions are negligible
after event selection, so that both the neutral-current Drell-Yan-like and triboson
processes have been omitted from our simulation.

Signal and background hard scattering events have been generated at the
leading-order accuracy with the {\sc MadGraph5}\_aMC@NLO program, using the
NNPDF~2.3 parton density sets, and matched with the parton shower infrastructure and
hadronization framework of {\sc Pythia 8.2}~\cite{Sjostrand:2014zea}. Our
simulation strategy furthermore follows the MLM merging
scheme~\cite{Mangano:2006rw} for combining event samples featuring a
different jet multiplicity. We have simulated the response of an LHC-like
detector by employing the {\sc Delphes~3.0} program~\cite{deFavereau:2013fsa} and finally
reconstructed all final-state jets by means of the anti-$k_T$
algorithm~\cite{Cacciari:2008gp} as embedded into
{\sc FastJet}~\cite{Cacciari:2011ma}.

The transverse momentum $p_T^\ell$ and pseudorapidity $\eta^\ell$ of all
electron and muon candidates are required to satisfy
\be
   p_T^\ell > 20~{\rm GeV} \qquad\text{and}\qquad
   |\eta_\ell| < 2.5\ ,
\ee
and we consider jet candidates whose transverse momentum $p_T^j$ and
pseudorapidity $\eta^j$ fulfill
\be
   p_T^j > 40~{\rm GeV} \qquad\text{and}\qquad |\eta_j| < 2.5\ .
\ee
We have moreover imposed that all reconstructed objects are isolated from each
other in the transverse plane, their angular distance $\Delta R$ being required
to be larger than 0.4 (0.5 in the case of two jets).
In order to optimize the selection to push the signal-to-noise ratio to a large
level, we use the {\sc MadAnalysis~5} software~\cite{Conte:2012fm} to implement
our phenomenological analysis. We require the presence of at least one
reconstructed lepton and at most three reconstructed jets,
\be
  N_\ell \geq 1 \quad\text{and}\quad N_j\leq 3 \ ,
\ee
and constrain the
amount of transverse missing energy to satisfy
\be
 \Emiss > 200 ~{\rm GeV}\ .
\ee
We moreover veto all events featuring either isolated photons with a $p_T$ greater than
10~GeV or $b$-tagged jets, using $b$-tagging efficiencies and misstagging rates
typical of the CMS detector (and implemented in the {\sc Delphes} detector
parameterization). The above selections allow us to reduce all backgrounds (the
$b$-tagged jet requirement specifically aiming to reduce the $t\bar{t}$ background)
to a large extent while maintaining a signal efficiency of about 50\%.

\begin{figure}
\includegraphics[width=3.0in]{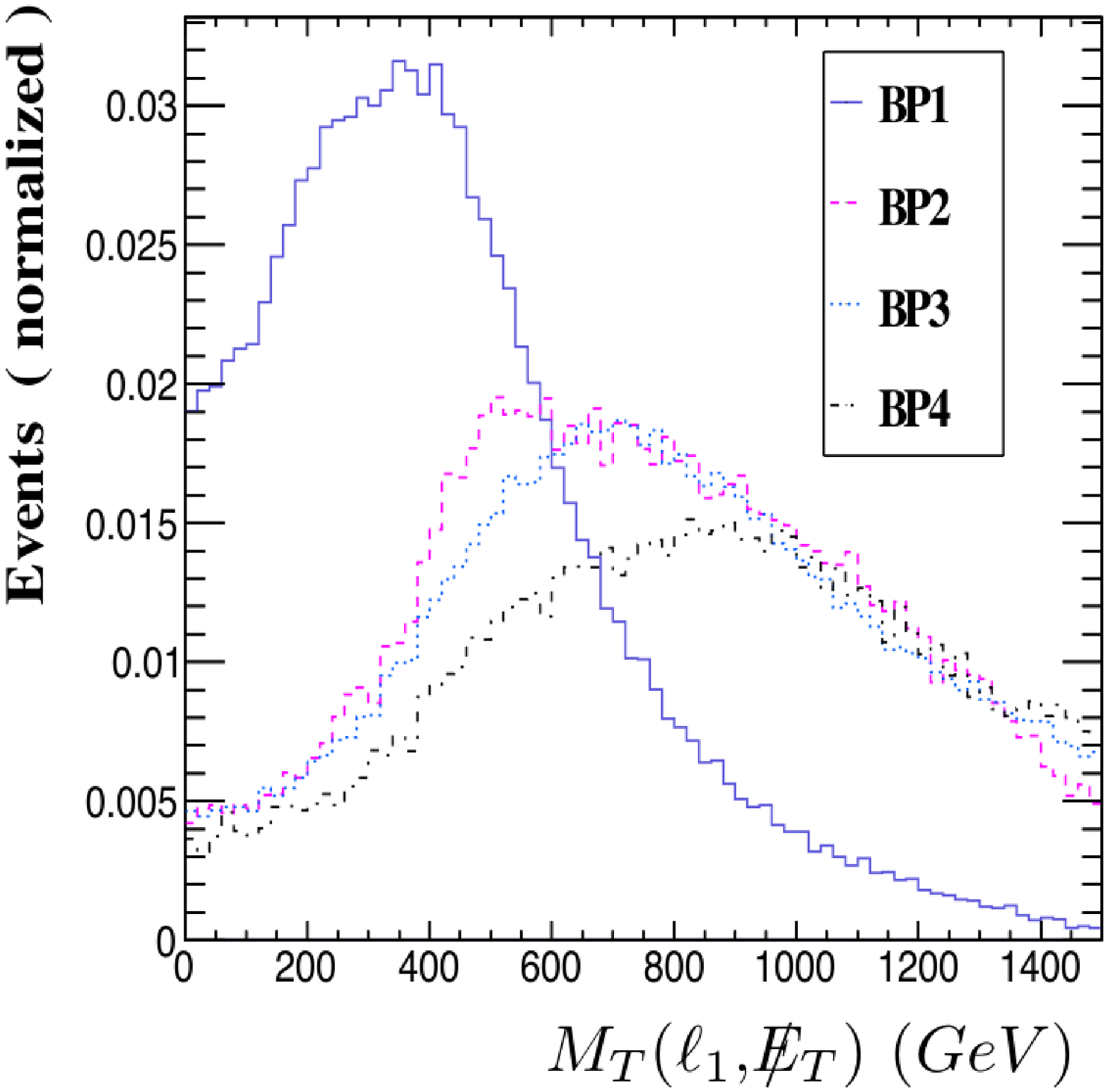}
\includegraphics[width=3.0in]{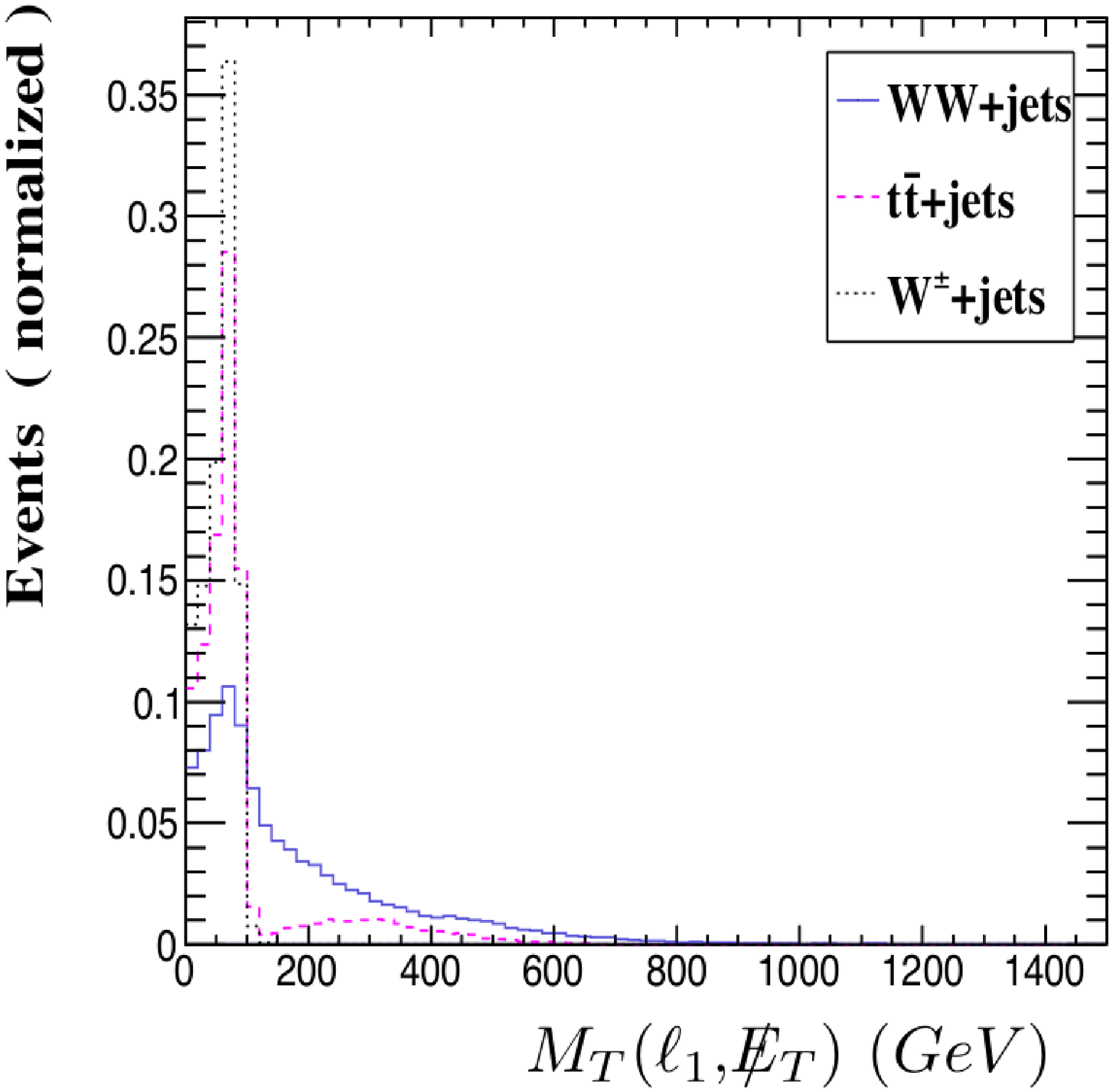}
\caption{\sl Distribution in the transverse mass $M_T$  of the
  leading lepton and the missing transverse momentum for the different benchmark
  point signals (left) and for the SM background (right). The distributions are
  shown after the basic acceptance selections defined in the text and normalized
  to 1.} \label{fig:mtl1}
\end{figure}

\begin{figure}
\includegraphics[width=3.0in]{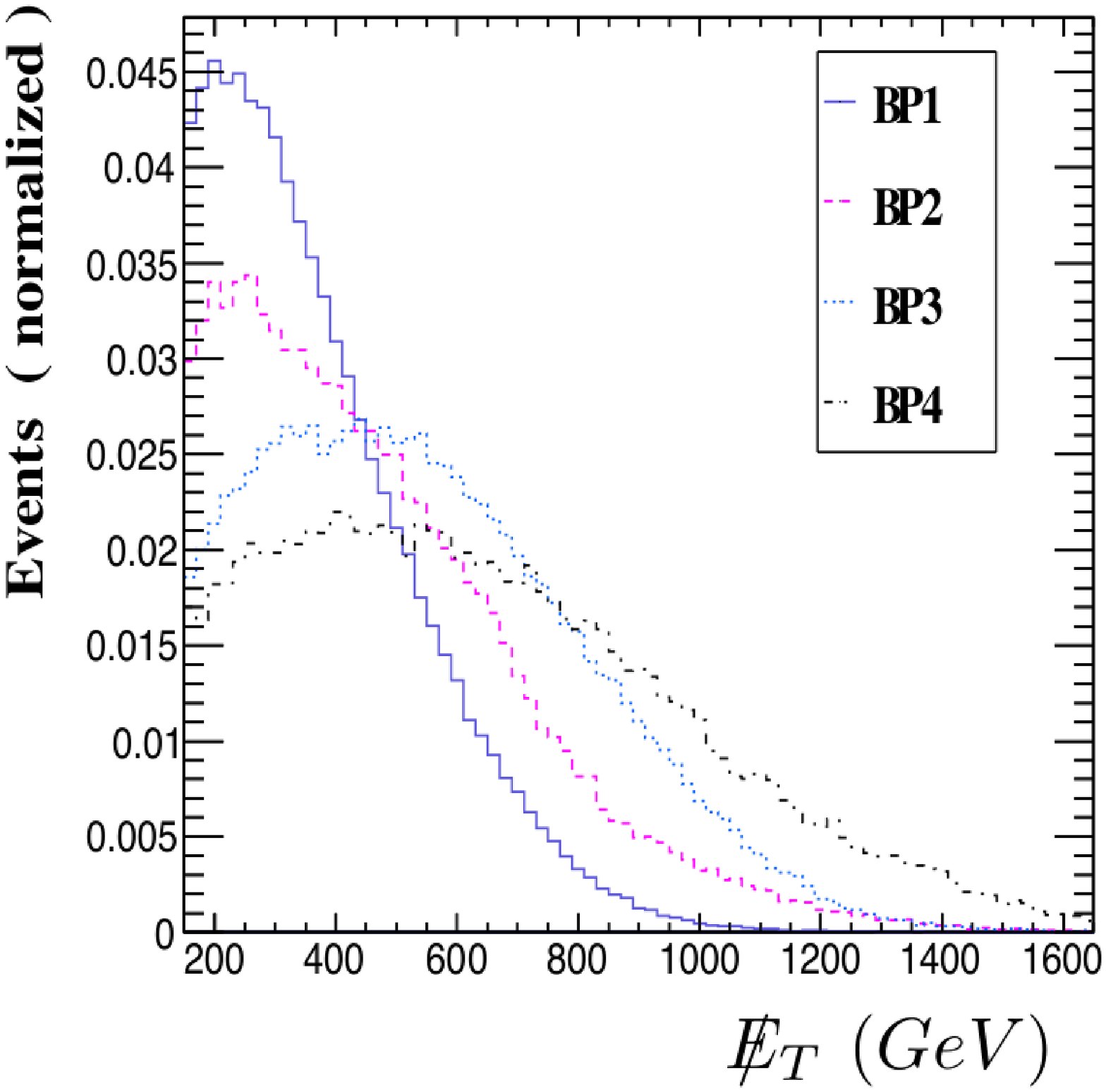}
\includegraphics[width=3.0in]{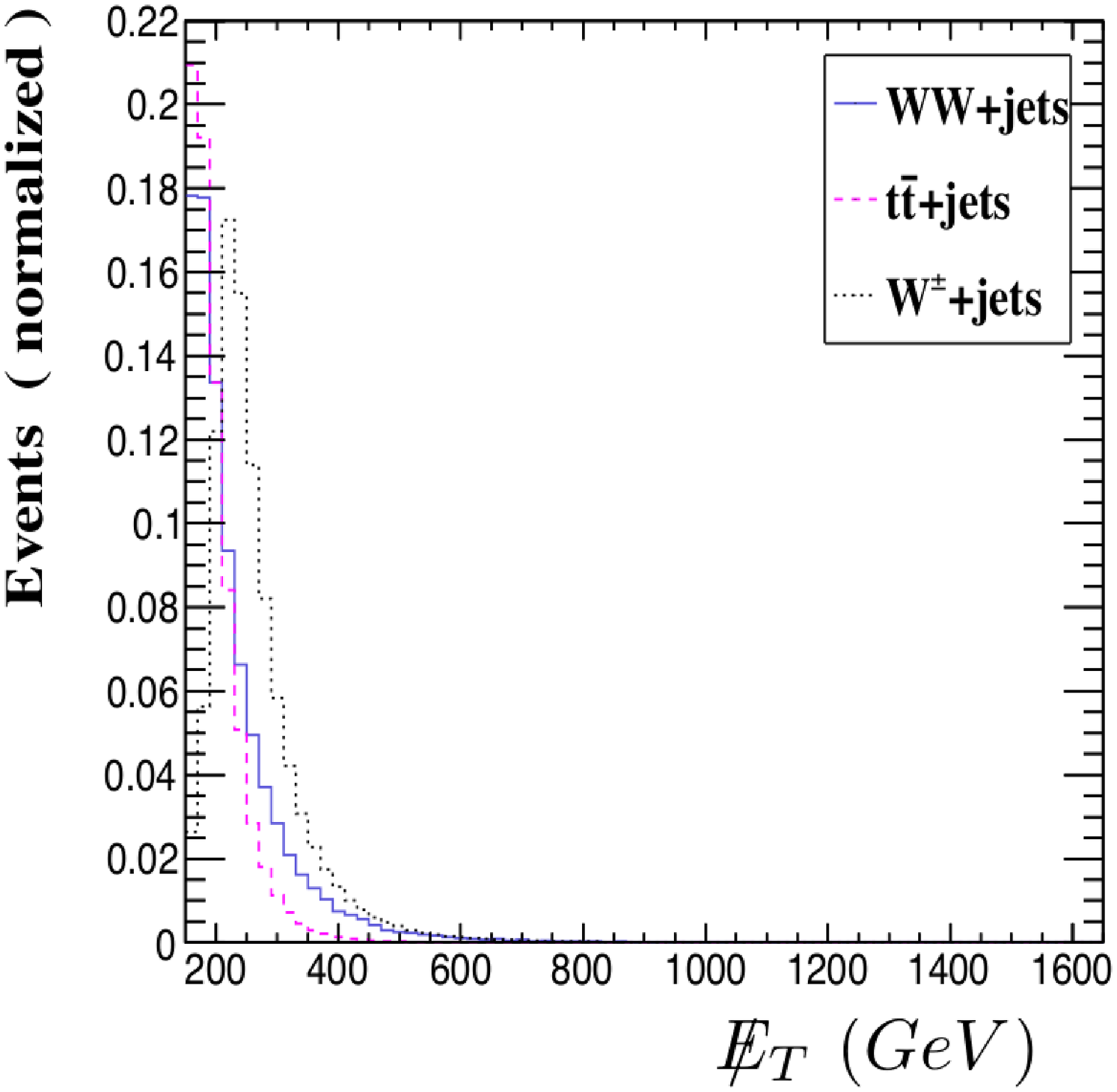}
\caption{\sl Same as in Fig.~\ref{fig:mtl1} but for the missing transverse
  energy $\Emiss$.} \label{fig:met}
\end{figure}

\begin{figure}
\includegraphics[width=3.0in]{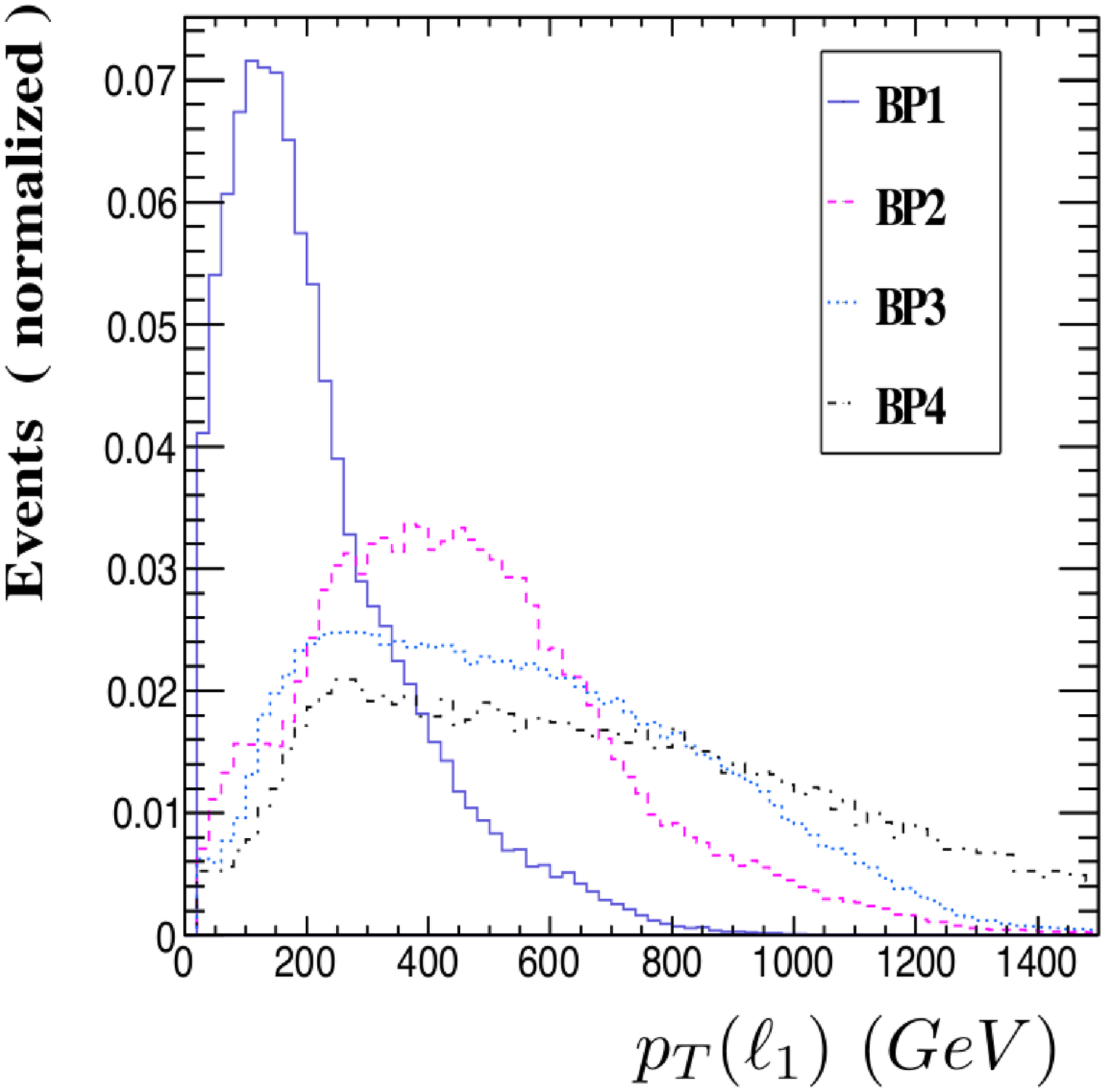}
\includegraphics[width=3.0in]{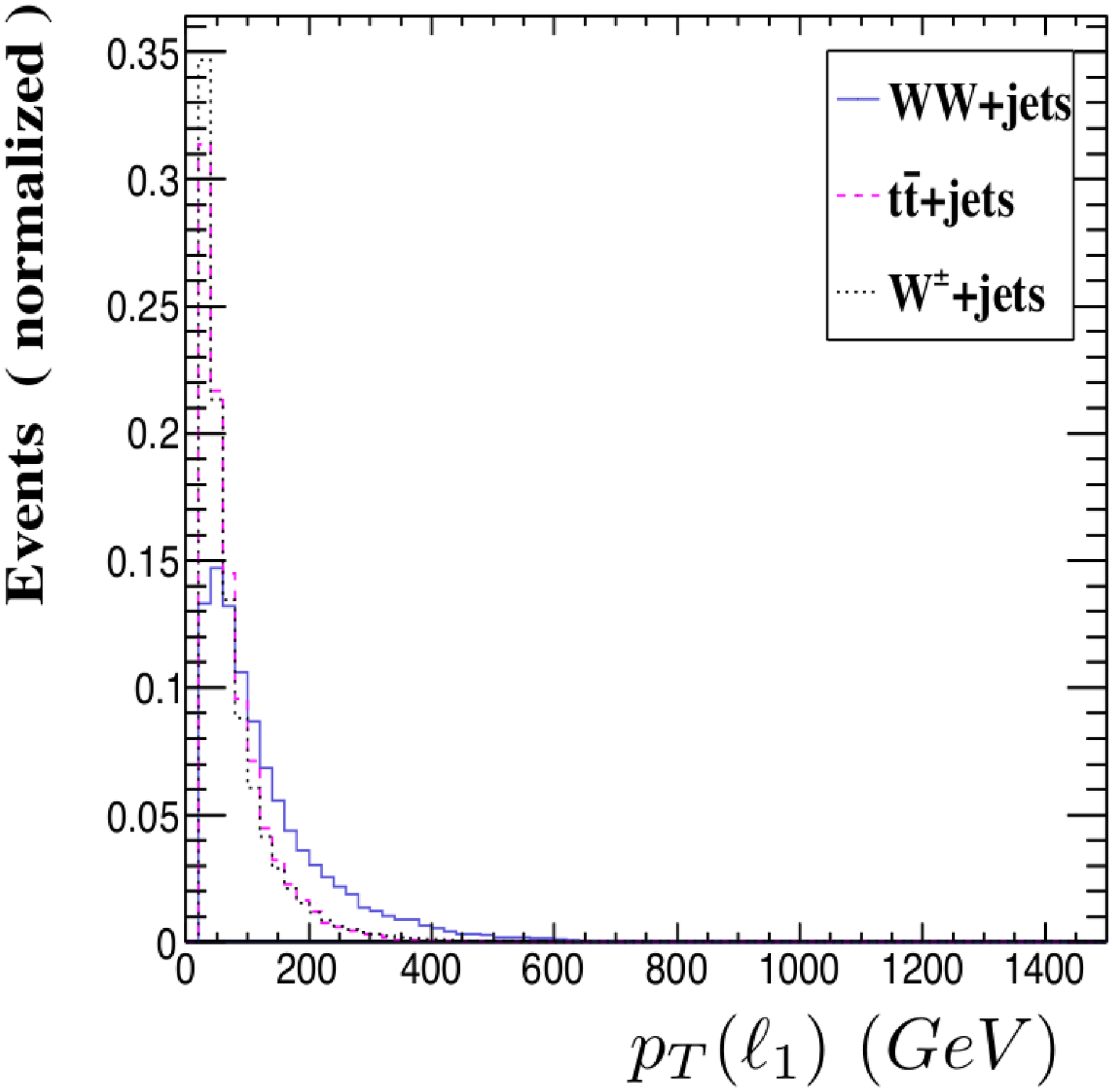}
\caption{\sl Same as in Fig.~\ref{fig:mtl1} but for the transverse momentum of
  the leading lepton.}
\label{fig:ptl1}
\end{figure}

The largest background contribution comes at this stage from charged-current
Drell-Yan-like events and still overwhelms the signal. The latter however features
leptons and missing energy originating from the decay of very massive
superpartners. This can
be used to design an appropriate selection, relying on, \eg, the transverse mass
$M_T(\ell_1,\Emiss)$ of the system made of the leading lepton and the missing
transverse energy. This observable is represented in Fig.~\ref{fig:mtl1} for
both the four signal benchmark scenarios (left) and the backgrounds (right)
after that all the previous selections have been applied. We additionally
compare the signal and background distributions in
the missing transverse momentum (Fig.~\ref{fig:met}) and in the transverse
momentum $p_T(\ell_1)$ of the leading lepton (Fig.~\ref{fig:ptl1}). This
demonstrates that all these three
observables yield neat handles for enhancing the signal over background
ratio, the signal spectra for each case being much harder than the background ones. Analyzing
the signal only, we observe that the distributions are harder for the
BP3 and BP4 scenarios than for the BP1 and BP2 scenarios. This results from the
superparticle spectrum associated with these scenarios (see
Table~\ref{tb:spectrum}), which features an important mass gap between the light
LSP and the charged sleptons that decay into the LSP and the corresponding
lepton. This contrasts with the BP1 and BP2 benchmarks where leptons also
originates from slepton decays, but where there is not such a large mass gap
with the LSP. From the above considerations, we impose a set of three selections,
\be
  \Emiss > 250~{\rm GeV}\ , \qquad
  M_T(\ell_1,\Emiss) > 250~{\rm GeV} \qquad\text{and}\qquad
  p_T(\ell_1) > 100~{\rm GeV} \qquad\text{for}\qquad \ell=e,\, \mu\ ,
\label{selection_cuts}\ee
which yields to an important background rejection, as illustrated in
Table~\ref{tb:cutflow1} for an integrated luminosity of 100~fb$^{-1}$.

\begin{table}
\begin{center}
\setlength\tabcolsep{4pt}
\renewcommand{\arraystretch}{1.4}
\begin{tabular}{|c | c c c c | c c c|}
  \hline
  & BP1 & BP2 & BP3 & BP4 & diboson & top-antitop & Drell-Yan\\
  \hline
  Preselection & 216 & 83 & 299 & 45 & 2065 & 7192 & $5.94\times 10^{5}$ \\
  $M_T(\ell_1,\Emiss) > 250~{\rm GeV}$  &
    153 & 77 & 279 & 42 & 521 & 708 & 142\\
  $p_T(\ell_1) > 100~{\rm GeV}$ &
    134 & 75 & 274 & 42 & 440 & 559 & 124\\
  $\Emiss > 250~{\rm GeV}$ &
    113 & 67 & 258 & 40 & 229 & 149 & 69 \\
  \hline
\end{tabular}
\end{center}
\caption{\sl Number of events surviving each step of our selection strategy for
  the four signal benchmark scenarios and the three main background components.
  We assume an integrated luminosity of 100~fb$^{-1}$.}
\label{tb:cutflow1}
\end{table}

The LHC turns out to be sensitive to the BP3 and BP1 scenarios that feature a
lighter $W_R$ boson with a mass of about 2.7~TeV with a
statistical significance of $11\sigma$ and $5\sigma$ respectively. Although the
BP2 and BP4 scenarios suffer from the reduction of the signal cross section
induced by the $W_R$-boson mass of about 3.5~TeV, significances of $3\sigma$ and
$1.86\sigma$ are obtained. As evident from Fig.~\ref{fig:ptl1}, the BP2, BP3 and
BP4 benchmarks implies a $p_T(\ell_1)$ distribution that is very hard, as
expected from the decay pattern of Eq.~\eqref{eq:bp234dec}, which provides an
extra way to increase the significance. For example, a selection of 200~GeV on
the leading slepton transverse momentum would suppress the total number of
background events to 250, for an integrated luminosity of 100~fb$^{-1}$,
whereas the number of signal events drops to 72, 63, 235 and 38 for the BP1,
BP2, BP3 and BP4 scenarios respectively. This improves the significance for BP4
to about $2.3\sigma$. Our analysis hence clearly suggests that LRSUSY sleptons
with masses of around 800~GeV could be accessible at the current run of the LHC
with a luminosity that is as low as 50 fb$^{-1}$, provided the $W_R$-boson mass
is around 3~TeV.

\begin{table}
\begin{center}
\setlength\tabcolsep{5pt}
\renewcommand{\arraystretch}{1.4}
\begin{tabular}{|c | c c c c | c c|}
  \hline
  & BP1 & BP2 & BP3 & BP4 & diboson & top-antitop \\
  \hline
  $n_\ell \geq 2$; $p_T(\ell_1) > 200~{\rm GeV}$ &
    55  & 21 &  77 & 14 & 94  & 50 \\
  $p_T(\ell_2) > 40~{\rm GeV}$ &
     50 & 18 &  66 & 13 & 72  & 38 \\
  $M_T(\ell_2,\Emiss) >  50~{\rm GeV}$  &
     46 & 17 &  63 & 13 & 41  & 21\\
  \hline
\end{tabular}
\end{center}
\caption{\sl Number of events surviving each step of a typical LRSUSY selection
  targetting the signature $(ii)$ of Eq.~\eqref{final_states}. We assume (and omit), as a
  preselection, the analysis depicted in Table~\ref{tb:cutflow1}.}
\label{tb:cutflow2}
\end{table}

We have verified that for the second signature of Eq.~\eqref{final_states},
there is a marked suppression in the number of selected signal events, so that
any potentially visible excess would require a significantly higher integrated
luminosity. The impact of a typical analysis strategy is shown in
Table~\ref{tb:cutflow2}, the selection requiring the presence of at least two
charged leptons. The second lepton is constrained to be harder
than 40~GeV and the transverse mass for this second lepton and
the missing momentum is required to be larger than 50~GeV.

\subsection{Collider consequences of the LSP nature in LRSUSY models}

For LRSUSY scenarios with a bino-dominated LSP, charged sleptons and sneutrinos
both decay into the neutralino LSP and either a lepton or a RH neutrino with a
large branching fraction. The RH neutrino then decays into an $\ell j j$ system
so that the full decay chain
is connected with a signature that includes two leptons, two
jets and missing transverse energy. In the case where sneutrinos cannot decay
into a $\nu_{R}\tchi^{0}_{1}$ final state, they instead decay invisibly to a
$\nu_{L}\tchi^{0}_{1}$ system which does not yield any multileptonic final state.
As a result of the decay tables of our benchmarks, decay modes exhibiting three or more leptons are rare and the corresponding
number of events amounts to about 10\% of the number of dilepton events.
The situation is slightly different for
scenarios where the LSP is a sneutrino. One expects lepton-enriched final
states, as intermediate charginos that can be produced in the longer decay
chains lead to additional leptons (like for the BP1 point).
In this case, the number of events featuring three leptons amounts to $20-30\%$ of
the number of expected dilepton events.
With a luminosity of $100$~fb$^{-1}$, this number is however too
small to get any statistically-signif\mbox{}icant way to
distinguish a sneutrino LSP scenario from the corresponding neutralino LSP scenario.
However, there is  hope for the high-luminosity phase of the LHC,
as for the study of the rarer same-sign dilepton signature that seems very
unlikely to  yield any visible signals at the low-luminosity phase of the LHC.

LRSUSY sneutrino LSP scenarios present also a very different phenomenology from the
corresponding MSSM scenarios where the MSSM is extended by RH neutrino
superfields. In this last case, the Lagrangian includes Dirac mass terms for the
neutrinos and the lightest stau is often close in mass to the sneutrino. Due to
the small associated Yukawa coupling, the lightest stau is
long-lived~\cite{Gupta:2007ui}, which contrasts with our scenarios where the
stau is much heavier than the sneutrinos (see Eqs.~\eqref{eq:sleptonmass},
\eqref{eq:sneutrino1} and \eqref{eq:imsnumass}). In LRSUSY, the
next-to-lightest superpartner (NLSP) turns thus out to be another
particle, and the presence of larger couplings of the sneutrino to the other
particles guarantees that the NLSP is typically not long-lived.

The LHC is expected 
 in general to be more sensitive to LRSUSY realizations with a sneutrino LSP than
for MSSM setups with a RH neutrino. Equivalently, higher
superpartner masses could be reached. In the MSSM, multileptonic final states
arise from electroweakino decays that can be either directly produced or
indirectly produced from squark and gluino decays~\cite{Arina:2013zca,%
Arina:2015uea,Banerjee:2016uyt}. The production rate of electroweakinos with
masses lying beyond $1$~TeV is however small, whereas the presence of the RH
gauge sector in LRSUSY enhances it, at least if the $W_R$-boson mass is not much
greater than 3~TeV. Corresponding events feature, in addition,  a larger amount of
missing transverse momentum and at least one very hard lepton, which helps to
suppress the SM background.

\section{Summary and conclusions}
\label{sec:conclusion}

In this work we investigated  the phenomenology of right-sneutrinos in LRSUSY models. The
RH neutrino superfields being part of a doublet, the expectations are
dif\mbox{}ferent from those of models with singlet neutrino superfields. We
studied the impact of having an LSP RH sneutrino and
shown that it could be a viable DM candidate satisfying all
cosmological constraints. In particular, the DM annihilation cross section does
not need any specific enhancement to accommodate the relic abundance observations. The only coupling responsible for
DM annihilation in the early Universe is a gauge coupling and there is a wide
range of sneutrino masses yielding a correct relic density.

We devised four benchmarks for our comparative study of LRSUSY scenarios
with a sneutrino and with a neutralino LSP, two benchmarks featuring a
right-sneutrino LSP, and two a neutralino LSP. We  considered two values of
the $W_R$-boson mass, chosen to agree with limits stemming from dijet
measurements at the LHC. In addition, the benchmarks have been imposed to
satisfy dark matter constraints, experimental mass limits and
the measurements of several low-energy observables.

We investigated the LHC phenomenology of our LRSUSY scenarios, focusing on
a LRSUSY signal originating from resonant slepton production via a $W_R$-boson
exchange and containing one or more charged leptons and missing transverse
momentum.
For optimistic scenarios with a light $W_R$ boson whose mass is 2.7~TeV, we have
shown that the transverse mass of the lepton/missing momentum systems, the missing
transverse energy and the transverse momenta of the final-state leptons are
suitable observables to differentiate the signal from the background even with a
low LHC luminosity of $100$~fb$^{-1}$, both for scenarios with a sneutrino
and with a neutralino LSP. For benchmarks with a heavier $W_R$ boson with a mass equal to
$3.5$~TeV, the discovery of the signal would require a slightly higher
integrated luminosity, which the LHC should be able to attain within a few years of
running.

Turning to signatures featuring more than two leptons, we have found that
the associated signal rates are lower than the dilepton one by about 20-30\%.
This contrasts with neutralino LSP scenarios where this number goes down to
about 10\%. Using these probes for
distinguishing different LRSUSY setups therefore requires the high-luminosity
phase of the LHC. Sneutrino LSP in left-right supersymmetry scenarios also reveal significant
differences with sneutrino LSP setups in supersymmetric realizations where the MSSM is
extended by a gauge-singlet neutrino superfield. The mass difference between the stau and
the sneutrino is larger in LRSUSY, and the stau here is in general neither the NLSP,
nor long-lived.

Considering right sneutrino as the DM candidate in LRSUSY realizations presents novel
features in the particle spectrum, the corresponding dark matter analysis
and the subsequent collider signals.
In particular, we propose that a resonantly-enhanced slepton production cross
section, otherwise overlooked in typical LRSUSY signals, allows for an improved sensitivity
to heavier slepton searches at the LHC. This also constitutes a promising
supersymmetric signal of physics beyond the MSSM.

\begin{acknowledgments}
The authors would like to thank Arindam Chatterjee for taking part to the
initial discussions of this work, Manuel Krauss for providing the LRSUSY
{\sc SPheno} code and Mihailo Backovic for his help with {\sc MadDM}.
MF thanks the NSERC for partial financial support under grant number SAP105354,
KH and HW acknowledge support from the H2020-MSCA-RICE-2014 grant number
645722 (NonMinimalHiggs), the work of SKR has been partially supported by
funding available from the Department of Atomic Energy, Government of India,
for the Regional Centre for Accelerator-based Particle Physics (RECAPP) at the
Harish-Chandra Research Institute, and BF has been supported in part by French
state funds managed by the Agence Nationale de la Recherche (ANR), in the
context of the LABEX ILP (ANR-11-IDEX-0004-02, ANR-10-LABX-63).
\end{acknowledgments}

  \end{document}